\documentclass[12pt]{article}

\usepackage[margin=1in]{geometry}
\usepackage[T1]{fontenc}
\usepackage{setspace}
\usepackage{caption}
\usepackage{graphicx}	
\usepackage{amsmath}	
\usepackage{amssymb}	
\usepackage{url}

\title{The link between star formation and gas in nearby galaxies}

\author{Robert Feldmann$^{1}$\footnote{robert.feldmann@uzh.ch}}
\date{{\small $^1$Institute for Computational Science, University of Zurich, Winterthurerstrasse 190, CH-8057 Zurich, Switzerland}}

\begin{document}

\maketitle

\begin{abstract}
Observations of the interstellar medium are key to deciphering the physical processes regulating star formation in galaxies. However, observational uncertainties and detection limits can bias the interpretation unless carefully modeled.
Here I re-analyze star formation rates and gas masses of a representative sample of nearby galaxies with the help of multi-dimensional Bayesian modeling. Typical star forming galaxies are found to lie in a `star forming plane' largely independent of their stellar mass. Their star formation activity is tightly correlated with the molecular and total gas content, while variations of the molecular-gas-to-star conversion efficiency are shown to be significantly smaller than previously reported.
These data-driven findings suggest that physical processes that modify the overall galactic gas content, such as gas accretion and outflows, regulate the star formation activity in typical nearby galaxies, while a change in efficiency triggered by, e.g., galaxy mergers or gas instabilities, may boost the activity of starbursts.
\end{abstract}

\section{Introduction}
\label{sect:Introduction}

Understanding how galaxies form their stars remains one of the major goals of galaxy theory\cite{Krumholz2014c}. Empirical relations that link star formation to galaxy properties have provided many clues to this cosmic puzzle. The discovery of a relatively tight relation between star formation rate (SFR) and stellar mass of galaxies\cite{Noeske2007d, Daddi2007a} showed that star formation proceeds in a similar fashion in most star forming galaxies but with a highly redshift dependent normalization. While the physical origin of this star forming sequence (SFS) is not yet fully understood, it is likely linked to the accretion of gas onto galaxies and the growth of their parent dark matter halos\cite{Dave2008, Lilly2013c, Feldmann2015}.

A more direct way of studying galactic star formation is by analyzing the interstellar medium (ISM) of galaxies\cite{Boselli2014a, Genzel2015, Saintonge2016, Tacconi2017}. Observationally, the surface density of star formation is well correlated with the surface density of molecular gas\cite{Bigiel2008}. The physical interpretation of this empirical correlation is that both star formation and molecular hydrogen formation require low gas temperatures and high densities and thus occur in co-spatial locations of the ISM\cite{Krumholz2011a}.

Constraining gas masses of galaxies is observationally challenging and subject to various biases and selection effects. Fortunately, recent observations of carbon-monoxide (CO) and 21cm line emission make it now possible to study the molecular and neutral gas content of representative samples of nearby galaxies\cite{Saintonge2017, Catinella2018} thus enabling a more comprehensive analysis of galactic star formation, the ISM composition, and the link to gas accretion.

A major conclusion reached by these studies was that star formation in galaxies does not simply scale with the mass of the molecular reservoir as suggested by previous analyses of the molecular Kennicutt-Schmidt relation but that the efficiency of converting molecular gas into stars varies with the offset from the SFS\cite{Saintonge2011g, Shetty2013, Tacconi2017, Tacconi2020}. However, selection effects pose a main challenge for this interpretation given that a large number of galaxies in these samples have line emission below the detection limit. Bayesian modeling offers a way to mitigate biases arising from such detection limits and other observational limitations\cite{Kelly2007, Robotham2015, Feldmann2019a}.

The present study employs a Bayesian approach to model the multi-dimensional distribution of SFRs, molecular gas, and neutral gas masses in a representative sample of nearby galaxies\cite{Saintonge2017, Catinella2018} while accounting for detection limits and observational uncertainties. The efficiency of star formation in typical star forming galaxies is found to be largely constant both along and across the SFS. In contrast, the star formation activity of starbursts may be boosted by a high efficiency. Overall, the SFRs and total gas masses of galaxies are shown to be strongly correlated, suggesting that galactic star formation is regulated by physical processes involving gas accretion and galactic outflows. Valuable information about the gas accretion histories of galaxies may thus be gleaned from accurately constraining the slopes of the SFS and the corresponding neutral and molecular gas sequences.

\section{Results} 

\subsection{The star formation, neutral gas, and molecular gas sequences}

\begin{figure}
\begin{tabular}{c}
\includegraphics[width=160mm]{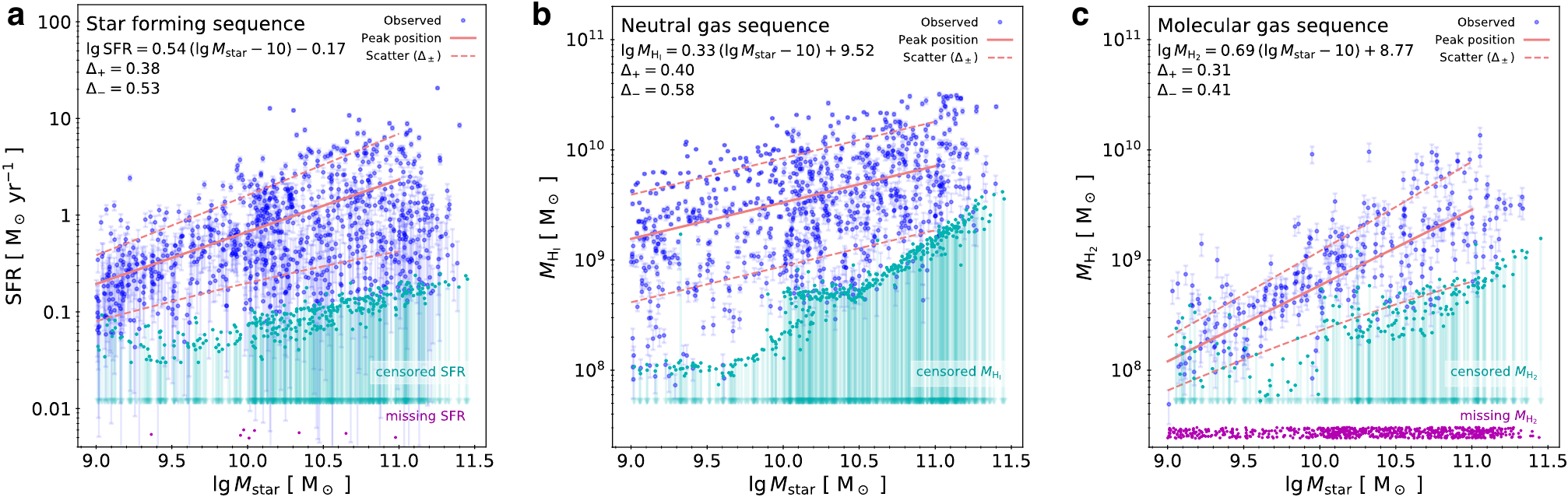}
\end{tabular}
\caption{{\bf Scaling relations of nearby galaxies.} 
Slope, normalization, and scatter of the star forming sequence ({\bf a}), neutral gas sequence ({\bf b}), and molecular gas sequence ({\bf c}). Points show the representative sample based on the xGASS / xCOLD GASS data sets\cite{Saintonge2017, Catinella2018}. Specifically, detected SFRs and gas masses are shown as blue circles with error bars indicating measurement uncertainties (one standard deviation). A large fraction of the observational data is either undetected/censored (cyan arrows) or missing (purple dots) necessitating careful modeling to avoid systematic biases. Peak position and scatter of each sequence, as determined by this study, are shown by solid and dashed lines. The peak position is defined as the mode of the conditional probability density of $\lg{}$SFR, $\lg{}M_{\rm H_I}$, and $\lg{}M_{\rm H_2}$ given $M_{\rm star}$. The predicted scaling of the peak position with stellar mass as well as the upward ($\Delta{}_+$) and downward ($\Delta{}_-$) scatter of each sequence for $M_{\rm star}=10^{10}$ $M_\odot$ galaxies are listed in the legend of each panel.}
\label{fig:fig1}
\end{figure}

Two different samples are used in the present analysis. First, a  `representative sample' of 1012 galaxies with stellar masses $9\leq{}\lg{}M_{\rm star}\leq{}11$ selected from the extended GALEX Arecibo SDSS Survey\cite{Catinella2018} (xGASS). Second, an extension of the representative sample (`extended sample') that includes 54 additional galaxies with molecular gas measurements from the CO Legacy Database for GASS\cite{Saintonge2017}  (xCOLD GASS) that are not in xGASS. Importantly, all galaxies within a given stellar mass range are included in the analysis, i.e., there is no ad hoc selection of galaxies according to their star formation activity. 

The joint distribution of SFRs, neutral gas, and molecular gas masses at fixed $M_{\rm star}$ is modeled as a non-Gaussian multivariate distribution with parameters that vary with $M_{\rm star}$ (see method section). This multi-dimensional distribution consists of a continuous component and a zero-component. The latter corresponds to galaxies with vanishing SFRs and gas masses while the former includes all other galaxies. The one-dimensional (marginal) distributions of SFRs and gas masses of the continuous component are modeled as a mixture of two gamma distributions. The first gamma distribution corresponds to SFRs or gas masses of ordinary star forming galaxies. A gamma distribution is adopted as it provides a better approximation to the distribution of SFRs at fixed stellar mass than a log-normal distribution\cite{Feldmann2017, Donnari2019}. The second, sub-dominant gamma distribution accounts for outliers with high SFRs (i.e, starbursts) or gas masses\cite{Sargent2012}.

The present study employs the Likelihood Estimation for Observational data with Python (LEO-Py) method\cite{Feldmann2019a} to compute the likelihood of the various distribution parameters taking into account the detection limits, missing entries, outliers, and correlations of the observational data (for either the representative or the extended sample). Starting from a weakly informative prior, the probability distribution of the distribution parameters is explored via a Markov Chain Monte Carlo (MCMC) method with the help of an affine-invariant ensemble sampler\cite{Foreman-Mackey2012a}. The mean parameter values obtained from the MCMC chain based on the representative (extended) sample define the fiducial (extended) model.

SFRs and gas masses of galaxies in the representative sample are shown in Fig.~\ref{fig:fig1}. Also shown are the peak position of the SFS (Fig.~\ref{fig:fig1}a), the neutral gas sequence (NGS, Fig.~\ref{fig:fig1}b), and the molecular gas sequence (MGS, Fig.~\ref{fig:fig1}c) as well as their scatter according to the fiducial model. These sequences refer to intrinsic galaxy properties because observational artifacts such as detection thresholds, missing values, and observational errors are accounted for in the multi-dimensional Bayesian modeling. A number of physical processes such as environmental effects\cite{Cortese2011, Bahe2015a}, fluctuations in the SFRs, or varying gas accretion rates\cite{Tacchella2016, Feldmann2019, Caplar2019a, Wang2019a} may be responsible for setting the normalization, slope, and scatter of these sequences. The parameters of the fiducial and extended models as well as the slopes and scatters of the SFS, NGS, and MGS are listed in Supplementary Tables 1-4 (see Supplementary Note 1).

The peak position of the SFS for a given stellar mass is defined as the mode of the $\lg{}{\rm SFR}$ distribution of typical galaxies (i.e., those belonging to the main gamma component of the model)\cite{Renzini2015, Feldmann2019a}. For gamma distributed SFRs, the peak position also corresponds to the average SFR. Analogous definitions are adopted for the NGS and MGS.

The SFS scales sub-linearly with a slope of 0.54 in qualitative agreement with previous results obtained with different approaches\cite{Speagle2014, Catinella2018}. The upward (downward) scatter for $M_{\rm star}\sim{}10^{10}$ $M_\odot$ galaxies is 0.38 dex (0.53 dex). The NGS has a much shallower slope (0.33) but a similar upward and downward scatter compared with the SFS. Among the three sequences, the MGS shows the steepest slope (0.69) and the lowest scatter (0.31 dex).

The lower scatter and steeper slope of the MGS compared with the SFS may suggest that the latter may be a consequence of the former. In this scenario, the SFS is a secondary relation created by the relatively tight correlation between $M_{\rm H_2}$ and $M_{\rm star}$ on one hand, and between SFR and $M_{\rm H_2}$ (the galaxy-integrated form of the molecular Kennicutt-Schmidt relation) on the other.

\subsection{The star forming plane}

\begin{figure}
\includegraphics[width=160mm]{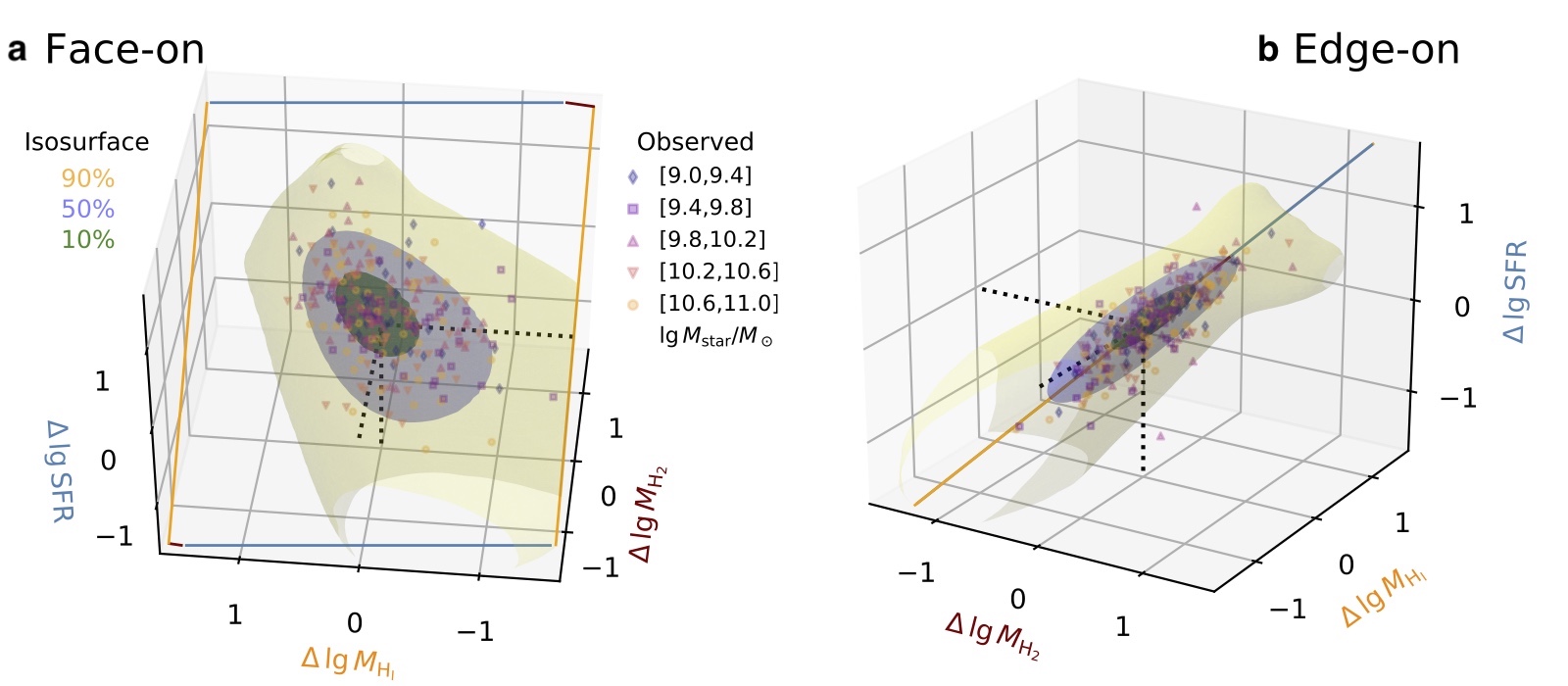}
\caption{{\bf  Star forming plane.} {\bf a} Face-on and {\bf b} edge-on view of the star forming plane. The star forming plane refers to the largely two-dimensional distribution of star formation rates (SFRs), neutral and molecular gas masses relative to the peak position of the star forming, neutral gas, and molecular gas sequence for a given stellar mass. Markers indicate the measured SFRs and gas masses of xGASS / xCOLD GASS observations with marker shapes and colors corresponding to different stellar masses (see legend). Regions bounded by the green, blue, and yellow isosurfaces include 10\%, 50\%, and 90\% of galaxies (without the zero component) according to the fiducial model. Solid lines mark the intersections of the star forming plane with the coordinate axes. The orientation of the star forming plane is calculated via a principal component analysis based on the probability density within the 50\% isosurface. The orientation of the star forming plane is only weakly dependent on stellar mass.}
\label{fig:fig2}
\end{figure}

The SFS, MGS, and NGS quantify how SFRs of galaxies and their gas masses scale with stellar mass but they provide limited information on how SFRs and gas masses are correlated. To study the latter, Fig.~\ref{fig:fig2} plots SFRs, $M_{\rm H_I}$, and $M_{\rm H_2}$ relative to the SFS, NGS, and MGS both for the observational xGASS / xCOLD GASS data and for the intrinsic properties as predicted by the fiducial model. Specifically, one of the axes shows $\Delta{}\lg{}{\rm SFR}= \lg\left({\rm SFR} / \lg{}{\rm SFR}_{\rm SFS}(M_{\rm star})\right)$, where ${\rm SFR}_{\rm SFS}(M_{\rm star})$ refers to the SFR of the SFS at stellar mass $M_{\rm star}$, see Fig.~\ref{fig:fig1}. The axes $\Delta{}\lg{}M_{\rm H_I}$ and $\Delta{}\lg{}M_{\rm H_2}$ are defined in an analogous fashion.

The surfaces shown Fig.~\ref{fig:fig2} are isosurfaces of probability density. They are calculated from a random sampling of the probability distribution of the fiducial model (with stellar masses drawn randomly from the representative xGASS / xCOLD GASS sample) via the marching cubes algorithm\cite{Lorensen1987}. The volumes enclosed by the isosurfaces contain 10\%, 50\%, and 90\% (from the innermost to the outermost isosurface) of the probability of the continuous component of the fiducial model. The isosurfaces are highly flattened in one direction. Fig.~\ref{fig:fig2}a,b show this `star forming plane' (SFP) in a face-on and edge-on view. The orientation of the SFP is calculated via a principal component analysis of all sample points within the 50\% isosurface. 

The orientation of the SFP could in principle depend on stellar mass. However, the present analysis suggests that such a dependence cannot be very strong. Fig.~\ref{fig:fig2} shows that the observed galaxies fall onto the star forming plane for all considered stellar masses. Furthermore, the orientation of the SFP as predicted by the fiducial model is also almost independent of stellar mass (see Supplementary Note 2). Hence, SFRs, $M_{\rm H_I}$, and $M_{\rm H_2}$, when measured relative to the peak position of their respective sequences, form an approximately 2-dimensional surface (the SFP) that is largely independent of stellar mass suggesting it is an approximately universal characteristic of (at least) nearby galaxies.

\begin{figure}
\begin{tabular}{c}
\includegraphics[width=135mm]{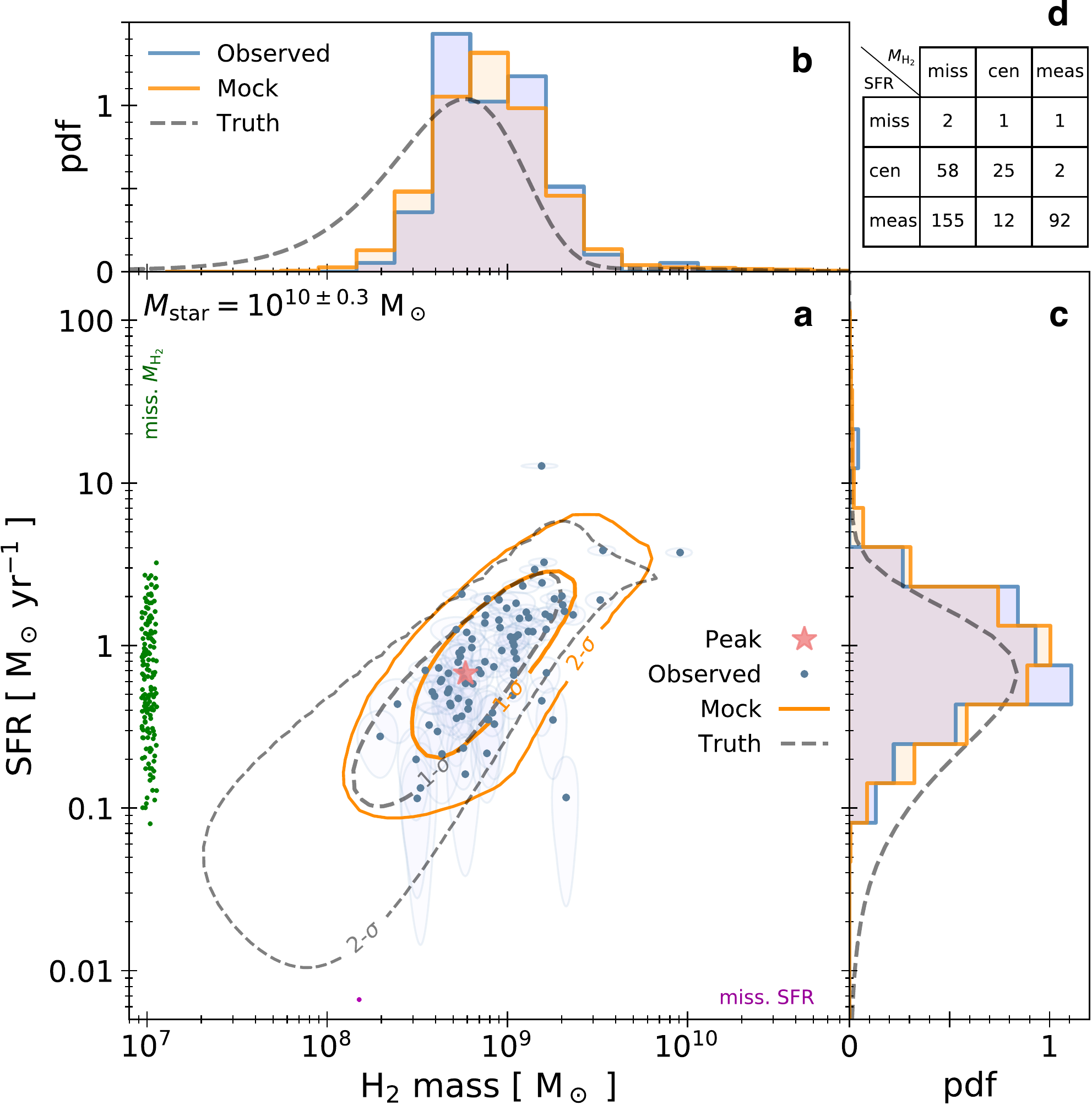}
\end{tabular}
\caption{{\bf Projection of the star forming plane along the $H_{\rm I}$ direction.} {\bf a} Distribution of star formation rate (SFR) and molecular gas mass ($M_{\rm H_2}$) in the representative xGASS / xCOLD GASS sample for galaxies with $M_{\rm star}=10^{10\pm{}0.3}$ $M_\odot$ (blue dots with 1-$\sigma$ uncertainty contours) and isocontours of the probability density of galaxies with the same stellar mass range according to the fiducial model (dashed lines include 68\% and 95\% of galaxies in the continuous component). Solid lines show the corresponding contours for a mock sample based on the fiducial model with observational errors and detection limits added. Observational data with missing SFRs (molecular gas masses) are shown as magenta (green) circles at the bottom (left) edge of panel {\bf a}. Censored observational data are omitted for clarity of presentation.
The red star indicates the peak position of the star forming sequence and the molecular gas sequence for galaxies with $M_{\rm star}=10^{10}$ $M_\odot$ according to the fiducial model. {\bf b}, {\bf c} Marginal probability density functions (pdf) of the xGASS / xCOLD GASS sample (blue histograms), the mock sample (yellow histograms), and the fiducial model (dashed curves) for $M_{\rm star}=10^{10\pm{}0.3}$ $M_\odot$ galaxies. {\bf d} Number of $M_{\rm star}=10^{10\pm{}0.3}$ $M_\odot$ galaxies in the representative xGASS / xCOLD GASS sample with measured ('meas'), missing ('miss'), and undetected ('cen') SFR and/or $M_{\rm H_2}$ values.
While the mock sample reproduces the observational data well, the differences between the mock data and the true model predictions suggest that significant biases can be introduced by censored, missing, and uncertain data.}
\label{fig:fig3}
\end{figure}

Fig.~\ref{fig:fig3} shows a projection of the 3-dimensional SFR, $M_{\rm H_I}$, and $M_{\rm H_2}$ space along the neutral gas direction for galaxies with $M_{\rm star}\sim{}10^{10}$ $M_\odot$. Given the narrow range of stellar masses, absolute SFRs and gas masses can be easily converted into quantities relative to their respective sequences and, hence, Fig.~\ref{fig:fig3} is a projection of the star forming plane onto the SFR -- $M_{\rm H_2}$ pair of axes. The SFR -- $M_{\rm H_2}$ diagram is close to an edge-on projection of the star forming plane, given its orientation shown in Fig.~\ref{fig:fig2}. Hence, this projection of the star forming plane corresponds to a tight relation between molecular gas mass and SFR, i.e., it is a galaxy-integrated version of the molecular Kennicutt-Schmidt relation\cite{Bigiel2008}. 

Fig.~\ref{fig:fig3} highlights two important points. First, the probability distributions of the observational data is well reproduced by the fiducial model after selection effects and observational uncertainties are taken into account. This shows that the underlying model provides a good description of the observational data. Secondly, there is a clear difference between the apparent (``mock'') and the actual (``true'') distribution of the model data thus highlighting the importance of properly modeling measurement uncertainties and data censoring in observational data. Here, data censoring refers to measurements that have been carried out but return values below a detection limit. The apparent relation between SFR and $M_{\rm H_2}$ is steeper than the actual relation as galaxies with low molecular masses are more likely to be censored than those with low SFRs. 

\subsection{Gas depletion times}
\label{sect:GasDepletion}

The slope of the SFR -- $M_{\rm H_2}$ relation is directly linked to the (molecular, neutral, total) gas depletion time, $t_{\rm dep}$, which is defined as the ratio between (molecular, neutral, total) gas mass and SFR. The total gas mass refers to the sum of molecular and neutral gas masses, and $t_{\rm dep}$ corresponds to the time it would take to convert the present gas reservoir into stars at the current SFR. The gas depletion time is a major parameter in galaxy models and its dependence on galaxy properties is an active area of observational and theoretical research\cite{Genzel2015, Tacconi2017, Janowiecki2018, Semenov2018}. 
Previous observational analyses\cite{Saintonge2011g, Boselli2014a, Tacconi2017, Tacconi2020} and numerical simulations\cite{Tacchella2016} have suggested that the molecular depletion time increases with a decreasing offset from the SFS, $\lg{}t_{\rm dep}\sim{}-0.5\times{} \Delta{}\lg{}{\rm SFR}$, for a broad range of offsets, stellar masses, and redshifts. If true, this result would suggest that star formation is not only regulated by the amount of molecular gas present but also by the molecular-gas-to-star conversion efficiency. The latter could arise from a variety of physical process operating in the ISM such as supersonic turbulence\cite{Krumholz2014c}. However, as pointed out above, the actual SFR -- $M_{\rm H_2}$ relation may differ from the apparent relation due to observational uncertainties and detection limits.

\begin{figure}
\begin{tabular}{c}
\includegraphics[width=140mm]{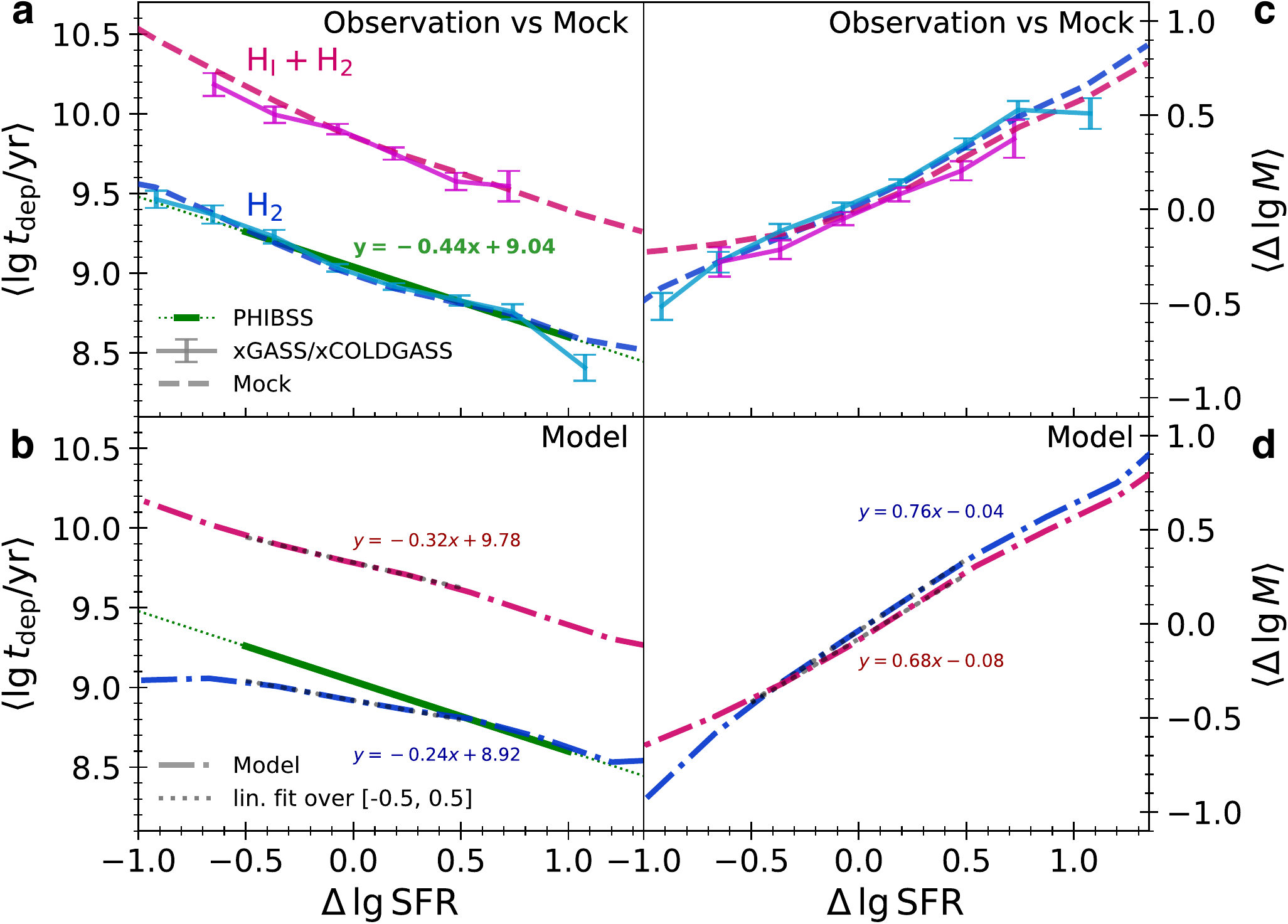}
\end{tabular}
\caption{{\bf Scaling of depletion times and gas masses.}
{\bf a} Average depletion times in the extended xGASS / xCOLD GASS sample of $M_{\rm star}\sim{}10^9-10^{11}$ $M_\odot$ nearby galaxies (solid lines with error bars) and in a mock sample with the same stellar mass distribution (dashed lines) showing good agreement. Blue/cyan lines (red/magenta lines) refer to molecular gas (to the sum of molecular and neutral gas including Helium). Galaxies with undetected or missing star formation rates (SFRs) are excluded from the analysis. Error bars correspond to standard errors of the bin averages. The solid green line shows the fit of the H$_2$ depletion time from the PHIBSS survey\cite{Tacconi2017} covering $z=0-4$ (dotted lines are extrapolations).
{\bf b} The scaling of the actual depletion times, i.e., if measured without observational errors and detection limits, for the galaxies in the mock sample. Galaxies with zero SFRs are excluded from the analysis. For typical offsets from the star forming sequence, the molecular gas depletion time shows only a mild dependence ($\propto{} {\rm SFR}^{-0.24}$) on SFR.
{\bf c}, {\bf d} Same as {\bf a}, {\bf b} but showing the average change in gas masses relative to the peak position of the corresponding gas sequence with offset from the star forming sequence. The peak position of the total gas sequence is given by the sum of the peak positions of the molecular and neutral gas sequences including Helium. Changes in star formation activity of typical starforming galaxies are tightly linked to changes in their molecular and total gas masses.}
\label{fig:fig4}
\end{figure}

Therefore, Fig.~\ref{fig:fig4} analyzes the molecular and total gas depletion times and their scaling with the offset from the SFS. Specifically, Fig.~\ref{fig:fig4}a shows the depletion times derived directly from the observational data as well as the depletion times in a mock sample based on the extended model (see section \ref{sect:Introduction}) after adding observational uncertainties and detection limits. The excellent agreement between observational and mock results suggests that the extended model well describes the observational data. The previously reported slope of $\sim{}-0.5$ based on galaxies with detected molecular gas masses is also recovered. However, the estimates of the depletion times and the calculated scalings are potentially biased as they do not properly account for missing and censored data.

Instead, Fig.~\ref{fig:fig4}b reports the actual scaling of the depletion times as predicted by the model. The scaling is significantly shallower ($-0.24$ for the molecular gas depletion time, $-0.32$ for the total gas depletion time) for typical offsets (-0.5 to 0.5) from the SFS. Hence, the gas depletion times are almost constant in normal star forming galaxies both along the SFS (given that the SFS and MGS have similar slopes\cite{Saintonge2016}, see Fig.~\ref{fig:fig1}) as well as across it (see also \cite{Scoville2016}). The scaling becomes steeper in galaxies (starbursts) that lie a factor of $\gtrsim{}3-5$ above the SFS indicating that the gas-to-star conversion efficiency is elevated in such systems as expected from studies of local ultra-luminous infrared galaxies\cite{Solomon1988}. 

Molecular and total gas masses vary with the offset from the SFS in a manner consistent with the results above, see Fig.~\ref{fig:fig4}c,d. In particular, the change of gas mass with the offset from the SFS becomes closer to linear once non-detected galaxies are included in the analysis. Again, this finding is consistent with a picture in which variations in the gas content, and not in the molecular-gas-to-star conversion efficiency, drive the star formation activity of typical (non-starbursting) nearby galaxies.

\begin{figure}
\includegraphics[width=140mm]{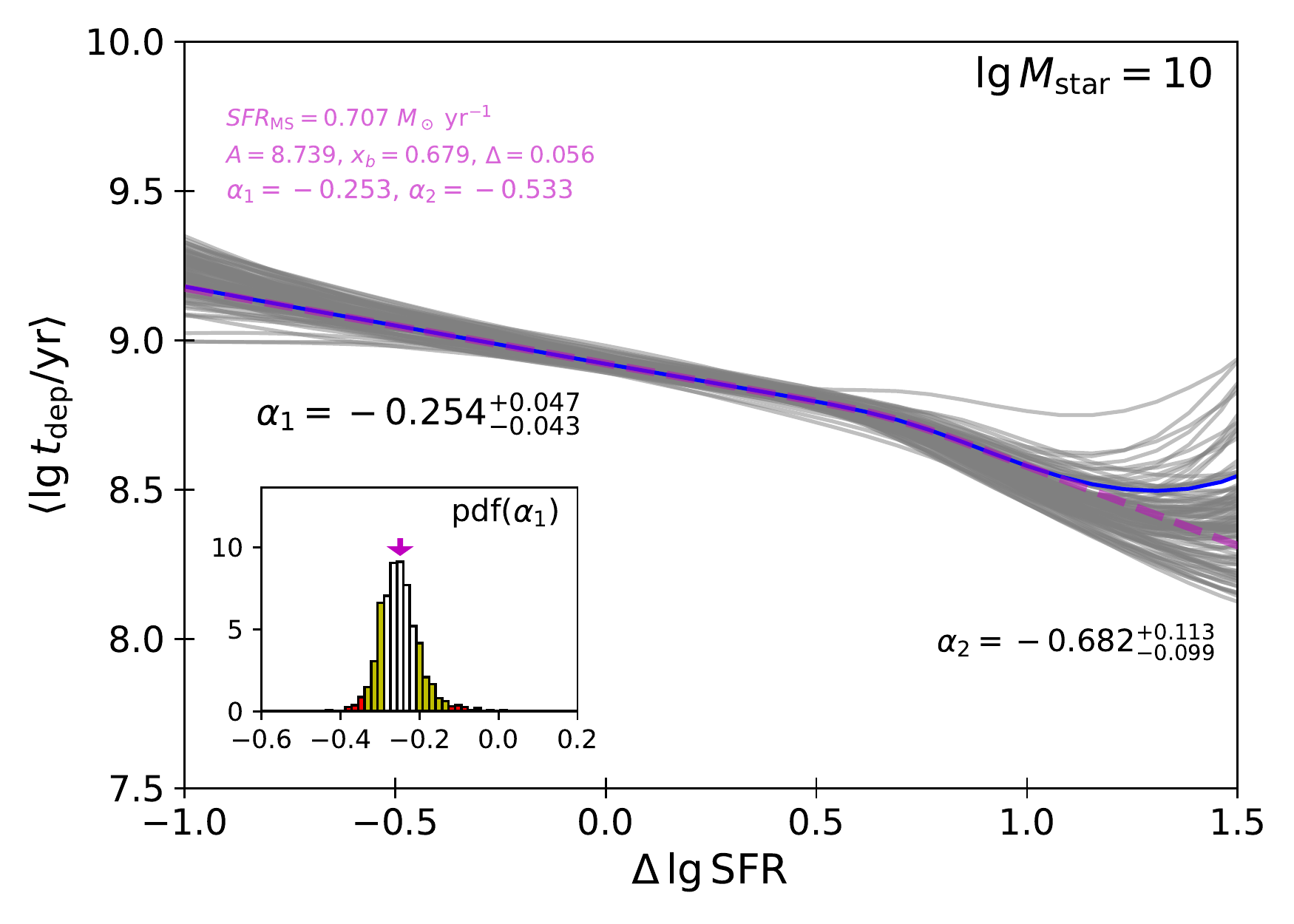}
\caption{{\bf Uncertainty of the molecular gas depletion time scaling.}
Scaling of the molecular gas depletion time with offset from the star forming sequence for $M_{\rm star}=10^{10}$ $M_\odot$ galaxies as predicted by the model trained  on the extended xGASS / xCOLD GASS sample and a quantification of the resulting uncertainty. The model prediction is shown by the blue solid line. A double-linear function provides a reasonable approximation, see the magenta dashed curve. The resulting fit parameters are reported in a magenta font in the top left corner of the main figure. Here, $\alpha_1$ is the slope of $\lg{}t_{\rm dep}$ with $\Delta{}\lg{}{\rm SFR}$ for non-starbursting galaxies, while $\alpha_2$ is the corresponding slope in the highly starforming regime.
Solid gray lines show the scaling of $t_{\rm dep}$ with $\Delta{}\lg{}{\rm SFR}$ for 100 different choices of the model parameters randomly selected from a Markov Chain Monte Carlo sampling. The inset panel shows the distribution of $\alpha_1$ based on 1000 such model parameters choices with the white (white + yellow) colored part of the histogram corresponding to the 68\% (95\%) credibility interval of $\alpha_1$. Median values and 16th to 84th percentile ranges are listed for both $\alpha_1$ and $\alpha_2$.}
\label{fig:fig5}
\end{figure}

Fig.~\ref{fig:fig5} quantifies the modeling uncertainty of the actual molecular gas depletion time, $t_{\rm dep, H_2}$, in galaxies with $M_{\rm star} = 10^{10}$ $M_\odot$.  The molecular gas depletion time is fit with a broken linear function between $\Delta{}\lg{}{\rm SFR}=-0.5$ and 1 for a large number of random draws of the model parameters from the MCMC chain. Specifically, $\langle{}\lg{}t_{\rm dep, H_2}/{\rm yr}\rangle{} = A + \alpha_1[x-x_{\rm b}] + [\alpha_2-\alpha_1]\ln\left([1 + \exp{\frac{x-x_{\rm b}}{\Delta{}}}]/2\right) \Delta{}$ is used as the fitting function, where $x=\Delta{}\lg{}{\rm SFR}$, $\alpha_1$ and $\alpha_2$ are the slopes for low and high values of $x$, $x_b$ and $\Delta{}$ are the break point and the smoothness of the transition from one slope to another, and $A$ is the value of $\langle{}\lg{}t_{\rm dep, H_2}/{\rm yr}\rangle{}$ at $x=x_b$. The median value of the slope of the molecular gas depletion time for non-starbursting galaxies ($\alpha_1$) with $M_{\rm star}\sim{}10^{10}$ $M_\odot$ is $-0.25$ and the 2.5th, 16th, 84th, and 97.5th percentiles are $-0.34$, $-0.30$, $-0.21$, and $-0.13$. Hence, the slope of the molecular gas depletion time (for non-starbursting galaxies) differs from zero (at the $2\sigma$ level)  but it is also significantly shallower than a -0.5 slope. Finally, the slope ($\alpha_2$) in the starbursting regime ($\Delta{}\lg{}{\rm SFR}\gtrsim{}0.6-0.7$) is steeper than $\alpha_1$ with $\alpha_2=-0.68^{+0.11}_{-0.10}$.

The analysis above implies that molecular gas depletion times of typical star forming galaxies depend only weakly on stellar mass and SFR. Specifically, combining the molecular depletion time scaling of galaxies that lie on the SFS and MGS (see Fig~\ref{fig:fig1}a,c and Supplementary Note 1) with the dependence of the depletion time on the offset from the SFS leads to
\begin{align}
\label{eq:tdepmol}
\begin{split}
t_{\rm dep, H_2}(M_{\rm star}, {\rm SFR}) 
& = \frac{M_{\rm H_2, SFS}}{{\rm SFR}_{\rm SFS}}(M_{\rm star}) \left(\frac{{\rm SFR}}{{\rm SFR}_{\rm SFS}(M_{\rm star})}\right)^{\alpha_1} \\
&= 0.87\,{\rm Gyr}\left(\frac{M_{\rm star}}{10^{10} M_\odot}\right)^{0.15} \left(\frac{{\rm SFR}}{{\rm SFR}_{\rm SFS}(M_{\rm star})}\right)^{-0.24},\,{\rm i.e.,}
\end{split}\\
\label{eq:tdepmol2}
t_{\rm dep, H_2}(M_{\rm star}, {\rm SFR}) & = 0.79\,{\rm Gyr}\left(\frac{M_{\rm star}}{10^{10} M_\odot}\right)^{0.28} \left(\frac{{\rm SFR}}{M_\odot\,{\rm yr}^{-1}}\right)^{-0.24}.
\end{align}

It is instructive to compare equation (\ref{eq:tdepmol}) with the result of a combined analysis of data sets spanning $z=0-4$\cite{Tacconi2017}. This latter study finds $t_{\rm dep, H_2}\propto{}M_{\rm star}^{0.09} ({\rm SFR}/{\rm SFR}_{\rm SFS}^z)^{-0.44} (1+z)^{-0.62}$, i.e., a steeper scaling with SFR and a dependence on redshift. 

Interestingly, the scaling $t_{\rm dep, H_2}\propto{}(1+z)^{-0.62}$ may be consistent with a molecular gas depletion time that has no explicit redshift dependence. This perhaps surprising result may be understood as follows. The normalization of the star forming sequence of galaxies increases quickly with redshift, ${\rm SFR}_{\rm SFS}^z=(1+z)^{2-3}{\rm SFR}_{\rm SFS}^{z=0}$ \cite{Lilly2013c, Whitaker2014b}, approximately in line with theoretical expectations from the scaling of the specific halo accretion rates \cite{Krumholz2012}. Consequently, if $t_{\rm dep, H_2}$ is independent of $z$ once $M_{\rm star}$ and SFR are given, then $t_{\rm dep, H_2}\propto{}M_{\rm star}^{0.15} \left({\rm SFR}/{\rm SFR}_{\rm SFS}^{z=0}\right)^{-0.24}=M_{\rm star}^{0.15} \left({\rm SFR}/{\rm SFR}_{\rm SFS}^z\right)^{-0.24}(1+z)^{-0.24\times{}2.5}\propto{}(1+z)^{-0.6}$. As a simple corollary, the molecular gas mass $M_{\rm H_2}$ of galaxies will also be a function of $M_{\rm star}$ and SFR alone, i.e., have no explicit dependence on $z$,
\begin{equation}
\label{eq:MH2}
M_{\rm H_2}(M_{\rm star}, {\rm SFR}) = 7.9\times{}10^{8}\,M_\odot\,\left(\frac{M_{\rm star}}{10^{10} M_\odot}\right)^{0.28} \left(\frac{{\rm SFR}}{M_\odot\,{\rm yr}^{-1}}\right)^{0.76}.
\end{equation}

The suggestion above is reminiscent of the fundamental metallicity relation\cite{Mannucci2010, Curti2020} which similarly explains the redshift evolution of the mass-metallicity relation\cite{Garnett2002, Tremonti2004a} with an underlying redshift-invariant dependence of the metallicity on both SFRs and $M_{\rm star}$. It is also similar to a proposed relation linking total gas mass fraction, stellar mass, and SFRs in a redshift independent manner\cite{Santini2014c}. Finally, given the $(1+z)^{2-3}$ scaling of the SFS, the redshift independence of equation (\ref{eq:tdepmol2}) is only in agreement with the scaling $t_{\rm dep, H_2}\propto{}(1+z)^{-0.62}$ if $\alpha_1$ is between $-0.31$ and $-0.21$.

\section{Discussion}
\label{sect:discussion}

The near constancy of $t_{\rm dep, H_2}$ in typical star forming galaxies suggests that their SFRs are largely driven by their molecular gas masses. The regulatory influence of physical processes that determine how efficiently molecular gas is converted into stars is thus limited, at least on global, galaxy-integrated scales in such galaxies. In contrast, a higher conversion efficiency appears to be the main driver of the excessively high star formation activity in starbursts. 

However, while galaxies near the star forming sequence have on average similar molecular gas depletion times, the ratio between $M_{\rm H_2}$ and SFR in any given galaxy can differ significantly from this average value as the model predicts a probability distribution, not a deterministic mapping, between gas mass and SFR. In particular, the scatter of SFRs at fixed molecular gas mass (and vice versa), see Fig.~\ref{fig:fig3}, may explain observations of galaxies with low SFR and, yet, significant amounts of molecular gas\cite{Suess2017}.

\begin{figure}
\includegraphics[width=160mm]{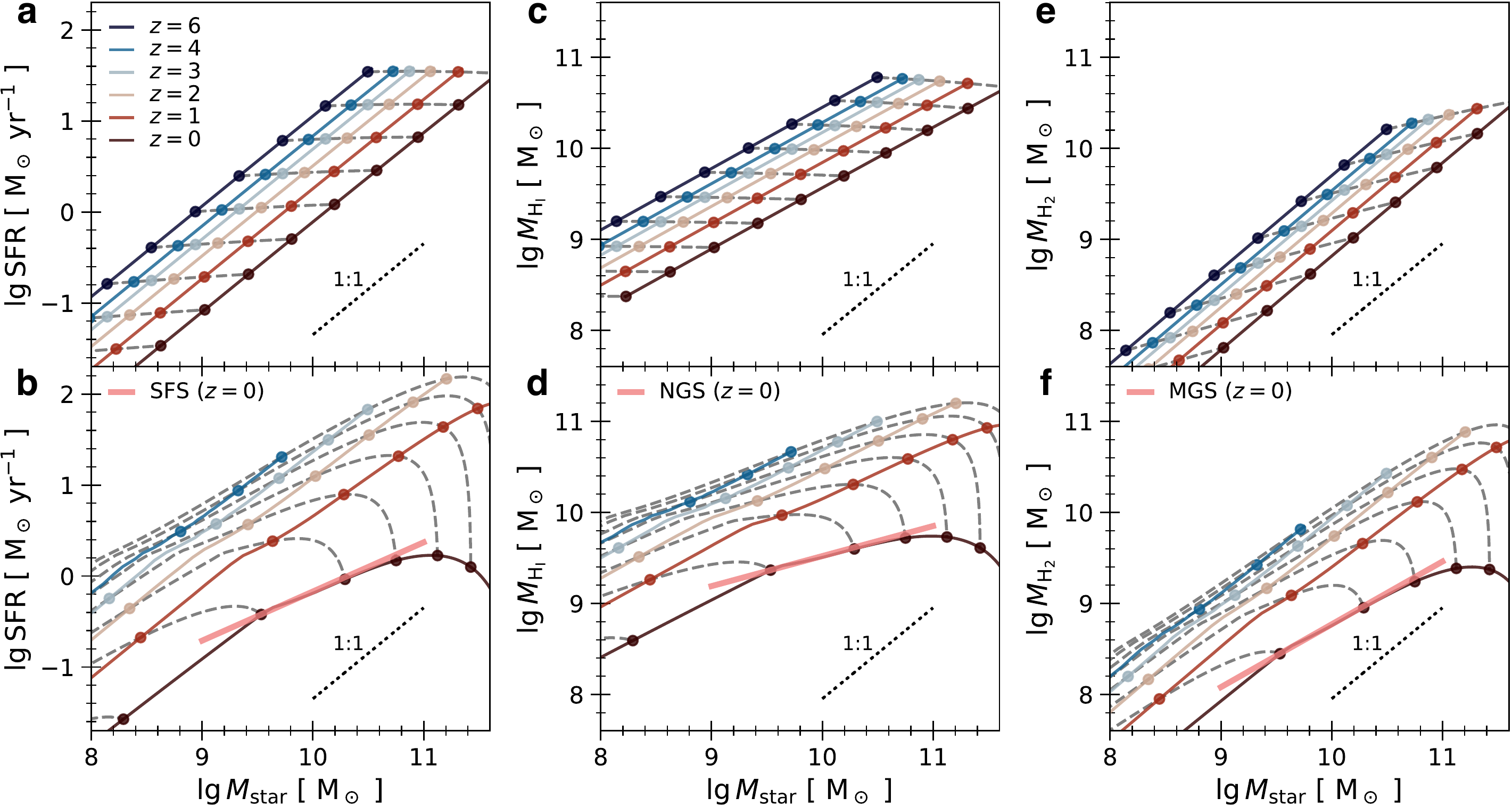}
\caption{{\bf Redshift evolution of galaxy scaling relations.} The star forming sequence (SFS, {\bf a}, {\bf b}), neutral gas sequence (NGS, {\bf c}, {\bf d}), and molecular gas sequence (MGS, {\bf e}, {\bf f}) and their redshift evolutions as predicted by two simple analytic models that link star formation rates (SFRs) and gas masses ($M_{\rm gas}$) of galaxies via ${\rm SFR} = M_{\rm gas} f_{\rm H2} / t_{\rm dep, H_2}$ and $t_{\rm dep, H_2}\propto{}M_{\rm star}^{0.28}\,{\rm SFR}^{-0.24}$. In both models, the stellar mass ($M_{\rm star}$) is the integral of the SFR, i.e., stellar mass loss and mergers are ignored. Furthermore, the molecular-to-total gas mass ratio ($f_{\rm H2}$) is assumed to depend only on stellar mass with $f_{\rm H2}(M_{\rm star})$ given by the scalings of the molecular and neutral gas sequences. In each panel, solid lines connect galaxy populations at a fixed redshift ($z=6-0$ from top to bottom), while dashed lines show the time evolution of individual galaxies. Linear slopes are indicated by dotted lines. {\bf a}, {\bf c}, {\bf e} Predictions of an equilibrium model in which $M_{\rm gas}$ does not change with time. The SFS has a slope of 1, while the slope of the MGS (NGS) is slightly steeper (less steep) than linear. {\bf b}, {\bf d}, {\bf f} Predictions of a model with a time-dependent $M_{\rm gas}$ such that $M_{\rm gas}$ peaks at earlier times in more massive galaxies (`downsizing'). This second model is successful in reproducing the sub-linear slopes of the SFS, MGS, and NGS (thick straight lines). Furthermore, it predicts that the slope of the SFS becomes steeper and more linear at higher redshift in qualitative agreement with observations\protect\cite{Whitaker2014b, Tomczak2016, Schreiber2017}.}
\label{fig:fig6}
\end{figure}

Fig.~\ref{fig:fig4} demonstrates that the molecular depletion time and the total gas depletion time have a similar scaling behavior with  $\Delta{}\lg{}{\rm SFR}$. This implies that the molecular-to-neutral gas ratio, and thus the molecular fraction $f_{\rm H_2}=M_{\rm H_2}/M_{\rm gas}$, is approximately constant across the SFS, i.e., for galaxies of a given $M_{\rm star}$, even including starbursts (see also Supplementary Note 3). The molecular-to-neutral ratio increases with increasing $M_{\rm star}$, however, as evidenced by the steeper slope of the MGS compared with the NGS. The molecular gas mass in nearby galaxies is thus primarily a function of $M_{\rm star}$ (via its effect on $f_{\rm H2}$) and the total gas mass. 

These considerations suggest an evolutionary model (see section \ref{sect:methods} for more details) in which the average star formation activity and stellar mass growth of star forming galaxies is determined by the time evolution of the total gas mass.
\begin{equation}
\label{eq:ToyModel}
{\rm SFR}(t) = \frac{M_{\rm gas}(t)}{t_{\rm dep}(M_{\rm star}(t), {\rm SFR}(t), t)}
\end{equation}
In the following discussion, $f_{\rm H_2}$ and $t_{\rm dep, H_2}$, and thus $t_{\rm dep}=t_{\rm dep, H_2}/f_{\rm H_2}$, are calculated from the empirically derived SFS, NGS, and MGS (see Fig.~\ref{fig:fig1}), with the additional scaling $t_{\rm dep, H_2}\sim{}{\rm SFR}^{-0.24}$ introduced in the previous section. An alternative version of ansatz (\ref{eq:ToyModel}) based on the reciprocal molecular gas depletion time is discussed in section \ref{sect:methods}.
Two specific choices for $M_{\rm gas}(t)$ are analyzed in more detail below. 

The case of an approximately constant gas mass, as predicted by a class of equilibrium galaxy formation models\cite{Bouche2010, Dave2012a}, provides a first example. In this case, the SFS is linear at all redshifts, while galaxies evolve along much more gradual trajectories (SFR $\sim{}$constant) in $M_{\rm star}-{\rm SFR}$ space, see Fig. \ref{fig:fig6}a. The predicted slopes of the MGS and NGS are slightly steeper (less steep) than linear, see Fig. \ref{fig:fig6}c,e. In either case, the predictions of this analytic model are in disagreement with the strongly sub-linear slopes of the SFS, NGS, and MGS shown in Fig.~\ref{fig:fig1}.

Perhaps surprisingly, the slope of the SFS will still be linear, even if the gas masses evolve with time, as long as the ratio of gas masses between galaxies is time-independent  and $t_{\rm dep}$ is a power-law function of $M_{\rm star}$ (see Supplementary Discussion). The empirical finding of a strongly sub-linear slope of the SFS thus suggests that gas mass histories of different galaxies are not scaled versions of each other. 

A second analytic model illustrates this result, see Fig. \ref{fig:fig6}b,d,f. In this model, the gas mass follows the typical growth histories of dark matter halos but is multiplied by additional factors that result in a downsizing effect of the gas mass, i.e., the gas mass reaches its maximum value at higher redshifts in more massive galaxies and then declines faster\cite{Santini2014c}. Not only does this second model reproduce the sub-linear slopes of the SFS, NGS, and MGS, it also results in a mass-dependent suppression of star formation at late times (quenching) and, furthermore, it predicts a steepening in the slopes of the scaling relations at higher redshift in qualitative agreement with observations \cite{Whitaker2014b, Tomczak2016, Schreiber2017}. More generally, the predicted slopes of the SFS, NGS, and MGS approach the corresponding predictions of the first (`equilibrium') model as the redshift increases.

The model described by equation (\ref{eq:ToyModel}) links the gas mass of galaxies to their star formation rates and stellar masses. The discussion above thus points to a picture in which physical processes affecting $M_{\rm gas}$ via gas inflows and outflows, such as cosmological gas accretion, hot gas cooling, a galactic fountain, and feedback from stars and black holes regulate the star formation activity and mass growth of typical, nearby galaxies\cite{Dekel2004, Keres2005, Hopkins2014, Vogelsberger2014, Schaye2015, Hobbs2020}. In contrast, the higher SFRs of today's starbursts appear to result from a higher efficiency of converting molecular gas into stars\cite{Genzel2010} and are thus likely related to changes in the physical state of the ISM on molecular clouds scales triggered by, e.g., galaxy mergers\cite{Hopkins2013c} or gas instabilities\cite{Dekel2014}. 

The quantitative results of this study are potentially subject to modeling choices and systematics inherent in the observational data sets. For instance, adopting lognormal instead of gamma distributions when modeling the SFRs and gas masses of galaxies increases the scaling coefficient $\alpha_1$ of the molecular gas depletion time from -0.25 to -0.22. In addition, the predicted slopes of the SFS, NGS, and MGS change by up to $\sim{}0.1$. However, the results of this paper are not qualitative affected by these changes. For example, in either case, the MGS (NGS) is predicted to be the sequence with the steepest (shallowest) slope and the smallest (largest) scatter. Secondly, to enable a fair comparison with the literature, the present analysis uses the xGASS and xCOLD GASS data as is. Hence, the accuracy of the model predictions may suffer from limitations related to observational systematics, such as those arising from the adopted conversion factors, flux aperture corrections, beam-size matching, and SFR calibrations.

Finally, the results presented here are based on measurements of nearby galaxies. Observations with the Atacama Large Millimeter/submillimeter Array, and other observatories, have begun to probe the ISM of high redshift galaxies in CO, CII, and continuum dust emission\cite{Walter2016, Tacconi2017, Fevre2019, Scoville2016}. Furthermore, observational challenges, such as the uncertain mapping of observables to physical properties\cite{Popping2019, Liang2019} and the large selection bias of most high-$z$ samples, can often be mitigated, e.g., by studying galaxy properties via multiple techniques and by surveying representative samples of high redshift galaxies\cite{Walter2016}. Additionally, complementary observations at radio wavelengths will soon constrain both obscured and unobscured SFRs down to a few $M_{\odot}$ yr$^{-1}$ up to $z=2$ \cite{Mancuso2015} and probe the ${\rm H_I}$ content of galaxies out to similar redshifts \cite{Blyth2015}.

Given the prospect of large representative samples of high redshift galaxies in the near future, it will be especially important to continue the development of methods to combine observations from multiple redshifts, observatories, and physical sources in a robust and reliable manner while accounting for detection limits, observational uncertainties, missing data, and data correlations. Indeed, these techniques will likely be critical to accurately quantify the link between gas properties, star formation rates, and stellar mass of galaxies across cosmic history, thus highlighting the increasing importance of statistics and data science in the study of galaxies.

\section{Methods}
\label{sect:methods}

\subsection{Observational data set.}
\label{sect:ObsDataSet}

The observational data is drawn from two related galaxy catalogs. The first is the  `representative sample' of the xGASS survey\cite{Catinella2018} (see \url{http://xgass.icrar.org}) which lists stellar masses, SFRs, and $H_{\rm I}$ masses (among other properties) of 1179 nearby galaxies ($0.01<z<0.05$) with a wide range of stellar masses ($M_{\rm star}=10^9-10^{11.5}$ $M_\odot$). The second catalog is the xCOLD GASS survey\cite{Saintonge2017} (see \url{http://www.star.ucl.ac.uk/xCOLDGASS}) which includes stellar masses, SFRs, and $H_2$ masses of 532 galaxies with the same redshift and stellar mass distribution. The CO luminosity to H$_2$ mass conversion factor adopted by xCOLD GASS is derived from a radiative transfer analysis of multiphase ISM simulations coupled with empirical relations between CII and CO line emission, gas phase metallicity, and offset from the SFS\cite{Accurso2017}. The two catalogs were merged with an outer join based on the provided GASS catalog identifiers resulting in a combined data set of 1234 nearby galaxies. The overlap between the two catalogs is very high (only 55 of the galaxies in the xCOLD GASS sample are not part of xGASS) which makes the combined xGASS / xCOLD GASS catalog an excellent data set to study the correlations between SFRs, $H_{\rm I}$, and $H_2$ masses of nearby galaxies.
The stellar masses in the joint catalog were replaced with updated SDSS Data Release 7 (DR7) median mass estimates\cite{Abazajian2009} available at \url{https://home.strw.leidenuniv.nl/~jarle/SDSS}. The original and the updated stellar masses agree to better than 1\% for all but a dozen of galaxies. The updated SDSS DR7 data also provide stellar mass measurement uncertainties (which are $\sim{}0.08-0.1$ dex for over 80\% of galaxies). The joint catalog is available as Supplementary Data, see Supplementary Note 4.

The analysis in this paper makes use of two subsamples generated from the joint catalog. First, all 1012 galaxies with $M_{\rm star}=10^9-10^{11}$ $M_\odot$ from the representative sample are selected from the joint catalog to form the `representative xGASS / xCOLD GASS sample'. Secondly, all 1066 galaxies with $M_{\rm star}=10^9-10^{11}$ $M_\odot$ from the joint catalog form the `extended xGASS / xCOLD GASS sample'. Both samples are similar, but the latter includes 43 additional galaxies with measured ${\rm H_2}$ masses and 11 additional galaxies undetected in ${\rm H_2}$. The average SFR of these 54 additional sources is $\sim{}7.1$ $M_\odot$ yr$^{-1}$ which is almost a factor 5 higher than the average SFR in the representative sample, while the average stellar masses ($\lg{}M_{\rm star}/M_\odot\sim{}10.2$) are virtually identical. This shows that starbursts make up a large fraction of these additional sources. Hence, the extended sample allows to better constrain the properties of starbursting galaxies at the cost of biasing the proportions between starbursting and non-starbursting galaxies.

About 30\% of the galaxies in the combined data set have SFR estimates but lack a quantification of their uncertainties. Two options were considered. A first possibility is to mark the SFRs of all such galaxies as missing which results in a large fraction of the available SFR estimates being excluded from the analysis.  An alternative approach consists of imputing SFR uncertainties based on a regression analysis. Specifically, the SFR uncertainty can be fit as function of SFR and stellar mass for those galaxies with provided SFR uncertainties. The analysis as presented in the paper follows the second approach but no substantive differences were found when the first option is chosen instead. All SFR measurements are censored if the SFR is lower than its measurement uncertainty. Measurement uncertainties of undetected $H_I$ (CO) sources are set to 1/5 the 5-$\sigma$ (1/3 the 3-$\sigma$) detection limit given in the xGASS (xCOLD GASS) catalog.

\subsection{Multi-dimensional model of star formation and gas content.}
\label{sect:MultidimModel}

The joint distribution of actual SFRs, molecular gas, and neutral gas masses at fixed stellar mass $M_{\rm star}$ is modeled as a multivariate distribution consisting of a continuous component and a discrete `zero-component'. The zero-component accounts for galaxies with vanishing SFRs and gas masses, while the continuous component models all other galaxies including regular star forming galaxies and outliers with high SFRs and/or gas masses\cite{Eales2017a, Feldmann2017, Feldmann2019a}. Hence, the probability density is
\[
p({\rm SFR}, M_{\rm H_I}, M_{\rm H_2} \vert{} \theta, \pi_0) = \pi_{0}\,\delta({\rm SFR})\delta(M_{\rm H_I})\delta{}(M_{\rm H_2}) + (1 - \pi_{0}) p_{\rm cont}({\rm SFR}, M_{\rm H_I}, M_{\rm H_2} \vert{} \theta),
\]
where $\theta$ is the set of parameters describing the continuous component, while $\pi_0$ is the probability of a galaxy to belong to the zero component and $\delta{}$ is the Dirac delta function. Both $\theta$ and $\pi_0$ are functions of $M_{\rm star}$. In addition to this 2-component model, an 8-component model was explored. In the latter, galaxies can belong (or not belong) to a zero component for each of SFR, $M_{\rm H_I}, M_{\rm H_2}$, i.e., they can have vanishing SFRs but not vanishing gas masses and vice versa. Thus, in the 8-component model there are 7 (partial) zero components and one fully continuous component. However, a Bayesian analysis showed that only 2 of the 8 components contribute significantly to the total probability. These two components are the zero-component and the continuous component in the 2-component model. Consequently, the 2-component model was adopted as the default choice.

The continuous component of the joint distribution is modeled with the help of a Gaussian copula. This approach generalizes multivariate normal distributions to allow for arbitrary continuous marginal distributions. The correlation structure is fully captured by the 3 off-diagonal coefficients of a $3\times{}3$ correlation matrix $R$, while the marginal (1-dimensional) distributions are modeled as a mixture of two gamma distributions. The first gamma component corresponds to SFRs or gas masses of ordinary star forming galaxies. It is parametrized by a shape ($a_{\rm SF}$) and scale ($b_{\rm SF}$) parameter. The second, sub-dominant gamma component accounts for outliers with high SFRs (i.e, starbursts) or gas masses\cite{Sargent2012}. Its parameters are $a_{\rm SF, out}$, $b_{\rm SF, out}$. Here, the scale $b_{\rm SF, out}$ is measured relative to the peak of the SFS. The peak position of the SFS is naturally defined\cite{Feldmann2019a} as the mode of the $\lg{}{\rm SFR}$ distribution of  galaxies after excluding starbursts and the zero component. For gamma distributed SFRs with parameters  $a_{\rm SF}$ and $b_{\rm SF}$, the peak of the SFS is at $a_{\rm SF}b_{\rm SF}$. The peak position is defined similarly for the NGS and the MGS. The fraction of the second gamma component in the gamma-mixture is given by $f_{\rm SF, out}$. The marginal distributions of $H_I$ and $H_2$ at fixed $M_{\rm star}$ are modeled in completely analogous fashion.

The slope and scale parameters of the primary gamma component are modeled as linear functions of $\lg{}M_{\rm star}$ with slopes $m$ and intercepts $n$ for each parameter, see Supplementary Note 1 for details. The slope angles ($\phi=\arctan(m)$) and perpendicular distances to the origin ($d=n\cos(\phi)$) are used as the actual model parameters\cite{Hogg2010} instead of $m$ and $n$. Given the relatively small number of galaxies with extreme SFRs or gas masses in the observational sample, no attempt is made in modeling the stellar mass dependence of $a_{\rm SF, out}$, $b_{\rm SF, out}$, and $f_{\rm SF, out}$. In contrast, a significant fraction of galaxies belongs to the zero component according to the predictions of the fiducial model. This fraction should depend on $M_{\rm star}$ given the increase in the quiescent fraction of galaxies with stellar mass\cite{Baldry2006}. Hence, the logit of $\pi_0$, defined as ${\rm logit}\,\pi_0 = \ln{}(\pi_0/(1-\pi_0))$, is modeled as a linear function of $\lg{}M_{\rm star}$, with slope angle ($\phi_0$) and perpendicular distance to the origin ($d_0$) as the main parameters.

The total number of parameters of the model is 26. There are $7\times{}3$ parameters that specify the slope and intercept of the stellar mass dependent parameters of the gamma-mixture for SFRs, neutral, and molecular gas masses, 3 correlation coefficients, and 2 parameters that define the stellar mass dependence of the zero-component. Estimates for all model parameters are provided in Supplementary Note 1.

\subsection{Bayesian analysis.}
The likelihood of the model parameters given the observational data is computed with LEO-Py\cite{Feldmann2019a}, available at \url{https://github.com/rfeldmann/leopy}. The likelihood estimate accounts for the observational uncertainties and detection limits of SFR and gas mass measurements. Measurement errors are assumed to be normally distributed with zero mean and a standard deviation given by the measurement uncertainty. Missing SFRs, $H_{\rm I}$, or $H_2$ masses are assumed to be missing at random (MAR), i.e., the probability that a given entry is missing may depend on other galaxy properties (e.g., on the stellar mass) but not on the missing value itself. 
Very weak priors are adopted for all model parameters. Uniform, bounded priors are used for each slope angle $\phi$ and perpendicular distance $d$. The prior for the 3-vector of the correlation coefficients is modeled as uniform over the sub-volume of $(-1, 1)^3$ for which the correlation matrix is positive semi-definite and zero otherwise.
The probability of model parameters given the observational data is given (modulo a constant of proportionality) by the product of the likelihood and the prior. However, since all adopted priors are uniform within the parameter bounds, this probability equals the likelihood (modulo a constant of proportionality) whenever all parameters are within their bounds, and 0 otherwise, thus simplifying the analysis.
The posterior probability distribution of the model parameters was sampled with the Markov Chain Monte Carlo (MCMC) ensemble sampler emcee\cite{Foreman-Mackey2012a}. Emcee was run for a total of 15000 steps using 1720 walkers and with a proposal scale parameter of 1.5. The first 4000 steps were considered burn-ins and discarded from the analysis. To reduce the wall-clock time, measurement uncertainties of stellar masses ($\sim{}0.09$ dex) were ignored. However, this simplification does not affect the presented results in a significant way, see Supplementary Tables 1-4. Furthermore, all MCMC calculations were run in parallel with MPI on 864 cores at the Swiss National Supercomputing Centre.

\subsection{Mock observations.}
\label{sect:Mock}
The present work uses mock data to confirm that the model provides a reasonable description of the observations and to construct the probability distribution of both actual and apparent galaxy properties for a given set of model parameters. The procedure below produces a mock catalog of specified size ($N_{\rm mock}$). First, $N_{\rm mock}$ stellar masses are drawn from the actual mass distribution of the xGASS / xCOLD GASS data set. Secondly, a given mock object is randomly assigned to either the zero component or the continuous component of the joint distribution with probability $\pi_0$ that depends on stellar mass. Mock objects in the zero component are assigned zero actual SFRs and gas masses.

For each mock object in the continuous component, a 3-dimensional random variate $\vec{z}=(z_{\rm H_2}, z_{\rm H_I}, z_{\rm SF})$ is drawn from a joint normal distribution with a covariance matrix given by a correlation matrix $R$. $R$ is fully specified by the model parameters.
Subsequently, $\vec{z}$ is converted into a 3-vector $\vec{y}$ of actual $M_{\rm H_2}$, $M_{\rm H_I}$, and SFR values via the mapping $y_X = F_X^{-1}\circ{}\Phi(z_X)$ where $X\in{}\{{\rm H_2}, {\rm H_I}, {\rm SF}\}$ corresponds to one of the observables ($M_{\rm H_2}$,  $M_{\rm H_I}$, or SFR), $F_X$ is the cumulative distribution of the observable corresponding to $X$ for a given $M_{\rm star}$, and $\Phi$ is the cumulative distribution of the standard normal distribution.

Thirdly, observational uncertainties are calculated for all mock objects based on the values of $\vec{y}$ and $M_{\rm star}$. Analogous to the approach discussed in section \ref{sect:ObsDataSet}, observational uncertainties of SFRs, $M_{\rm H_I}$, and $M_{\rm H_2}$ are estimated via linear regression using the value of these observables and $\lg{}M_{\rm star}$ as predictors. Observational errors $\delta{}\vec{y}$ (drawn from a standard multivariate normal distribution but rescaled such that the standard deviations are given by the previously calculated observational uncertainties) are added to $\vec{y}$ to obtain apparent (mock) observations, i.e., $\vec{y}^{\rm mock} = \vec{y} + \delta{}\vec{y}$. Finally, mock observations that fall below their respective detection limits (3-$\sigma$ for $M_{\rm H_2}$, 5-$\sigma$ for $M_{\rm H_I}$, 1-$\sigma$ for SFRs)  are marked as censored.

\subsection{Evolutionary Model}
\label{sect:EvolutionaryModel}

The paper introduces an analytic model of the form
\begin{equation}
\label{eq:EvModel}
{\rm SFR}(t, s) = \frac{M_{\rm gas}(t, s)}{t_{\rm dep}(M_{\rm star}(t, s), {\rm SFR}(t, s), t)} = \frac{f_{\rm H_2}(M_{\rm star}(t, s), {\rm SFR}(t, s), t) M_{\rm gas}(t, s)}{t_{\rm dep, H_2}(M_{\rm star}(t, s), {\rm SFR}(t, s), t)}
\end{equation}
and analyzes some of its predictions. 
In the equation above, $t$ is the cosmic time, $t_{\rm dep} = t_{\rm dep, H_2}/f_{\rm H_2}$ is the total gas depletion time, $f_{\rm H_2}=M_{\rm H_2}/M_{\rm gas}$ is the molecular gas fraction, $M_{\rm gas}(t, s)$ is a family of known gas mass histories, and $s$ is a one-dimensional parameter indicating a given evolutionary track. The SFR is the time derivate of the stellar mass, i.e., ${\rm SFR}(t, s) = \partial{}M_{\rm star}(t, s)/\partial{}t$, as long as stellar mass loss and mass accretion via galaxy mergers are ignored. The former can be partially accounted for by adopting the instantaneous recycling approximation\cite{Schmidt1963, Tinsley1980}, while the latter is a reasonable assumption given that star-forming galaxies acquire most of their stellar mass via in-situ star formation\cite{Behroozi2018a}.

As presented in section \ref{sect:GasDepletion}, the molecular gas depletion for typical star forming galaxies is a power-law function of $M_{\rm star}$ and ${\rm SFR}$ and potentially independent of $z$, i.e., $t_{\rm dep, H_2}(M_{\rm star}, {\rm SFR}) \propto{} M_{\rm star}^{\beta} {\rm SFR}^{\alpha}$. Furthermore, as discussed in section \ref{sect:discussion} and shown in Supplementary Figure 6, the molecular gas fraction depends on $M_{\rm star}$ (and potentially $t$) but not significantly on SFR. Hence, equation (\ref{eq:EvModel}) can also be written as
\begin{equation}
\label{eq:EvModel2}
{\rm SFR}(t, s) = a \left[M_{\rm star}(t, s)\right]^{-\beta/(1+\alpha)} \left[f_{\rm H_2}(M_{\rm star}(t, s), t) M_{\rm gas}(t, s)\right]^{1/(1+\alpha)}.
\end{equation}

Equation (\ref{eq:EvModel2}) together with $M_{\rm star}(0, s)=0$ is an initial value problem for any given fixed $s$. It can be solved numerically, e.g., with the \texttt{solve\_ivp} function from the Python \texttt{scipy.integrate} module, to obtain $M_{\rm star}(t, s)$ for all $t$. Subsequently, SFRs can be obtained from equation (\ref{eq:EvModel2}), molecular gas masses via $M_{\rm H_2}=t_{\rm dep, H_2} {\rm SFR}$, and neutral gas masses (including Helium) via $M_{\rm H_I}=M_{\rm gas}-M_{\rm H_2}$. As the evolutionary model uses the functional forms of the SFS, NGS, and MGS only indirectly, via $t_{\rm dep, H_2}$ and $f_{\rm H_2}$, it may not necessarily predict scaling relations in agreement with those shown in Fig.~\ref{fig:fig1}. For instance, the slope of their SFS will be exactly linear if galaxies evolve according to (\ref{eq:EvModel2}) with constant gas masses and $f_{\rm H_2}\propto{}M_{\rm star}^\gamma$ (see Supplementary Discussion). Comparing model predictions and observational measurements of the SFS, MGS, and NGS, thus allows to put constraints on the gas growth history of galaxies.

Equation (\ref{eq:tdepmol}) is a power-law approximation for $t_{\rm dep, H_2}$ as a function of SFR and $M_{\rm star}$. While this is the conventional choice, an alternative approach is to fit the reciprocal molecular depletion time $t^{-1}_{\rm dep, H_2}$ as a power-law function of $M_{\rm H_2}$ and $M_{\rm star}$, i.e.,
$t^{-1}_{\rm dep, H_2}(M_{\rm star}, M_{\rm H_2}) = a' M_{\rm star}^{-\beta'} M_{\rm H_2}^{-\alpha'}$. As shown in Supplementary Figure 8 (see Supplementary Note 5), $t^{-1}_{\rm dep, H_2}$ scales weakly with $M_{\rm H_2}$ ($\alpha'=-0.17$) in qualitative agreement with the weak SFR dependence of $t_{\rm dep, H_2}$ in equation (\ref{eq:tdepmol}). 
The SFRs of galaxies of a given $M_{\rm H_2}$ and $M_{\rm star}$ can be calculated with the help of $t^{-1}_{\rm dep, H_2}$ as follows: 
\begin{equation}
{\rm SFR} = t^{-1}_{\rm dep, H_2}(M_{\rm star}, M_{\rm H_2}, t) \, M_{\rm H_2} = a' M_{\rm star}^{-\beta'} M_{\rm H_2}^{1-\alpha'} ,\textrm{ with }M_{\rm H_2} = f_{\rm H_2}(M_{\rm star}, t) \, M_{\rm gas}.
\end{equation}
This alternative model is of the same form as equation (\ref{eq:EvModel2}) and thus can be solved in the same way. In fact, both models are identical if $\beta'=\beta/(1+\alpha)$, $\alpha'=\alpha/(1+\alpha)$, and $a'=a$.

\section*{Data Availability}
\setlength{\parskip}{4pt}
The xCOLD GASS\cite{Saintonge2017} and xGASS\cite{Catinella2018} catalogs are publicly available at \url{http://www.star.ucl.ac.uk/xCOLDGASS} and \url{http://xgass.icrar.org}. The combined xGASS / xCOLD GASS data set used in the present analysis is available as Supplementary Data, see Supplementary Note 4.

\section*{Code Availability}
\setlength{\parskip}{4pt}
LEO-Py\cite{Feldmann2019a} is publicly available at \url{https://github.com/rfeldmann/leopy}.
\vspace{0.3cm}

\section*{Acknowledgement}
The author thanks Reinhard Genzel, Simon Lilly, Lucio Mayer, and Romain Teyssier for insightful comments on the early draft of this manuscript. The author wishes to express his gratitude to Barbara Catinella for help with the xGASS data set. The author acknowledges financial support from the Swiss National Science Foundation (grant nos. 157591 and 194814). This work was supported by a grant from the Swiss National Supercomputing Centre (CSCS) under project IDs s926 and uzh18. This research has made use of NASA's Astrophysics Data System. 

The analysis presented in this work is partly based on data provided by the Sloan Digital Sky Survey (SDSS). Funding for SDSS and SDSS-II has been provided by the Alfred P. Sloan Foundation, the Participating Institutions, the National Science Foundation, the U.S. Department of Energy, the National Aeronautics and Space Administration, the Japanese Monbukagakusho, the Max Planck Society, and the Higher Education Funding Council for England. The SDSS Web Site is http://www.sdss.org/. The SDSS is managed by the Astrophysical Research Consortium for the Participating Institutions. The Participating Institutions are the American Museum of Natural History, Astrophysical Institute Potsdam, University of Basel, University of Cambridge, Case Western Reserve University, University of Chicago, Drexel University, Fermilab, the Institute for Advanced Study, the Japan Participation Group, Johns Hopkins University, the Joint Institute for Nuclear Astrophysics, the Kavli Institute for Particle Astrophysics and Cosmology, the Korean Scientist Group, the Chinese Academy of Sciences (LAMOST), Los Alamos National Laboratory, the Max-Planck-Institute for Astronomy (MPIA), the Max-Planck-Institute for Astrophysics (MPA), New Mexico State University, Ohio State University, University of Pittsburgh, University of Portsmouth, Princeton University, the United States Naval Observatory, and the University of Washington.
 
\section*{Author Contributions}
The author designed and carried out the project and wrote the manuscript.
 
\section*{Competing Interests}
The author declares no competing interests.

\section*{Correspondence}
Correspondence and requests for materials should be sent to robert.feldmann@uzh.ch.

\pagebreak

\begin{center}
  \textbf{\large The link between star formation and gas in nearby galaxies\\Supplementary Information}\\[.2cm]
  Robert Feldmann$^{1*}$\\[.1cm]
  {${}^1$Institute for Computational Science, University of Zurich, Winterthurerstrasse 190, \\CH-8057 Zurich, Switzerland}\\
  ${}^*$Electronic address: robert.feldmann@uzh.ch\\
\end{center}

\setcounter{equation}{0}
\setcounter{figure}{0}
\setcounter{table}{0}
\renewcommand{\theequation}{S\arabic{equation}}
\renewcommand{\figurename}{Supplementary Figure}
\renewcommand{\tablename}{Supplementary Table}

\section*{Supplementary Note 1}
\label{sect:SuppParameters}

The distribution of star formation rates (SFRs), neutral gas masses ($M_{\rm H_I}$), and molecular gas masses ($M_{\rm H_2}$) is modeled as a non-Gaussian multivariate distribution with  distribution parameters that may depend on stellar mass ($M_{\rm star}$), see section \ref{sect:MultidimModel}. The stellar mass dependence of a distribution parameter $\xi_X$ with $X\in\{\rm SF, H_I, H_2\}$ is encapsulated by a slope angle parameter $\phi_\xi^X$ and a perpendicular distance parameter $d_\xi^X$. Specifically, 
\begin{equation}
\label{eq:defparams}
g(\xi_X)(M_{\rm star}) = m_\xi^X \lg{}\frac{M_{\rm star}}{10^{10} M_\odot} + n_\xi^X, \,\,\,\textrm{ with } m_\xi^X = \tan \phi_\xi^X \textrm{ and } n_\xi^X = \frac{d_\xi^X}{\cos \phi_\xi^X}
\end{equation}
where $g$ is an appropriately chosen transformation of the parameter $\xi_X$. 

Supplementary Table~\ref{tab:ModelParameters} lists point estimates for the parameters of the fiducial model (see section \ref{sect:MultidimModel}) as well as percentiles of their 1-dimensional probability distributions obtained via Markov Chain Monte Carlo sampling. Supplementary Table~\ref{tab:ModelParametersExtended} contains the analogous parameter estimates for the extended model. For clarity, parameter names are labeled with the subscripts 'shape' instead of $a$, 'scale' instead of $b$, and 'frac' instead of $f$. The outlier distributions and the outlier fractions are assumed to not depend on stellar mass, i.e., the corresponding parameters have a slope angle parameter of zero which is thus not listed. The parameters $\rho_{\rm SF, H_I}$, $\rho_{\rm SF, H_2}$, and $\rho_{\rm H_I, H_2}$ are the correlation coefficients of the standardized variables $z_{\rm H_2}, z_{\rm H_I}, z_{\rm SF}$ (see section \ref{sect:Mock}), i.e., they are the off-diagonal entries of the correlation matrix of the Gaussian copula linking SFRs, molecular, and neutral gas masses. They are also assumed to be stellar mass independent. The transformation $g$ is the common logarithm for the shape and scale parameters of the gamma-mixture ($a_{\rm X}, b_{\rm X}, a_{\rm X, out}, b_{\rm X, out}$), it is the identity function for the outlier fraction ($f_{\rm X, out}$), and the ${\rm logit}$ function for $\pi_0$.  

Point estimates and percentiles of the slopes, normalizations, and scatter of the star forming sequence (SFS), neutral gas sequence (NGS), and molecular gas sequences (MGS) according to the fiducial and extended models are provided in Supplementary Tables~\ref{tab:DerivedParameters} and ~\ref{tab:DerivedParametersExtended}). Given that the peak of each sequence is defined as the mode of a gamma distribution, the peak position equals the product of the shape and scale parameters of this gamma distribution. The logarithm of these parameters scales linearly with $\lg{} M_{\rm star}$ with slopes $m^X_a$ and $m^X_b$ for $X\in\{\rm SF, H_I, H_2\}$, see equation (\ref{eq:defparams}). Therefore, the slope of each of the three sequences is given by $m^{\rm SFS} = m^{\rm SF}_a + m^{\rm SF}_b$, $m^{\rm NGS} = m^{\rm H_I}_a + m^{\rm H_I}_b$, and $m^{\rm MGS}=m^{\rm H_2}_a + m^{\rm H_2}_b$, respectively. Similarly, the normalization of each of the three sequences is given by
$n^{\rm SFS} = n^{\rm SF}_a + n^{\rm SF}_b$,  $n^{\rm NGS} = m^{\rm H_I}_a + n^{\rm H_I}_b$, and $n^{\rm MGS}=n^{\rm H_2}_a + n^{\rm H_2}_b$. The scatter (of the primary gamma component) of each sequence depends only on the shape parameter. Hence, for $M_{\rm star}=10^{10}$ $M_\odot$ galaxies it can be calculated directly from $n^{\rm SF}_a$, $n^{\rm H_I}_a$, and $n^{\rm H_2}_a$.

\begin{table}
\centering
\caption{{\bf Parameters of the fiducial model}}
\linespread{1.22}
\footnotesize
\begin{tabular}{l|rrrrrr}
parameter & mean & median & 16th perc. & 84th perc. & MAP & MAP$^{\rm ME}$ \\
\hline
$\phi_{\rm scale}^{\rm SF}$ & 0.661 & 0.662 & 0.623 & 0.698 & 0.663 & 0.674\\
$d_{\rm scale}^{\rm SF}$ & -0.126 & -0.124 & -0.161 & -0.0905 & -0.136 & -0.115\\
$\phi_{\rm shape}^{\rm SF}$ & -0.234 & -0.234 & -0.282 & -0.186 & -0.243 & -0.256\\
$d_{\rm shape}^{\rm SF}$ & -0.0126 & -0.0161 & -0.0609 & 0.0365 & 0.00131 & -0.0228\\
$d_{\rm scale, out}^{\rm SF}$ & 0.723 & 0.583 & 0.372 & 1.03 & 0.303 & 0.454\\
$d_{\rm shape, out}^{\rm SF}$ & -0.891 & -0.926 & -1.40 & -0.382 & -0.932 & -0.854\\
$d_{\rm frac, out}^{\rm SF}$ & 0.0610 & 0.0562 & 0.0229 & 0.0971 & 0.0765 & 0.0462\\
\hline
$\phi_{\rm scale}^{H_{\rm I}}$ & 0.323 & 0.323 & 0.282 & 0.364 & 0.321 & 0.322\\
$d_{\rm scale}^{H_{\rm I}}$ & 9.10 & 9.11 & 8.97 & 9.24 & 8.98 & 8.98\\
$\phi_{\rm shape}^{H_{\rm I}}$ & -0.00417 & -0.00403 & -0.0450 & 0.0367 & -0.00817 & -0.0154\\
$d_{\rm shape}^{H_{\rm I}}$ & -0.0768 & -0.0766 & -0.106 & -0.0472 & -0.0782 & -0.0784\\
$d_{\rm scale, out}^{H_{\rm I}}$ & 1.87 & 1.48 & 0.405 & 3.64 & 0.394 & 0.393\\
$d_{\rm shape, out}^{H_{\rm I}}$ & -0.945 & -1.19 & -1.81 & 0.256 & 0.464 & 0.472\\
$d_{\rm frac, out}^{H_{\rm I}}$ & 0.0235 & 0.00449 & 0.000848 & 0.0220 & 0.219 & 0.216\\
\hline
$\phi_{\rm scale}^{H_2}$ & 0.799 & 0.801 & 0.759 & 0.840 & 0.786 & 0.814\\
$d_{\rm scale}^{H_2}$ & 5.99 & 5.99 & 5.72 & 6.26 & 6.11 & 5.95\\
$\phi_{\rm shape}^{H_2}$ & -0.328 & -0.329 & -0.396 & -0.259 & -0.313 & -0.350\\
$d_{\rm shape}^{H_2}$ & 0.167 & 0.157 & 0.0887 & 0.251 & 0.101 & 0.102\\
$d_{\rm scale, out}^{H_2}$ & 0.833 & 0.683 & 0.412 & 1.21 & 0.957 & 1.18\\
$d_{\rm shape, out}^{H_2}$ & -0.966 & -1.02 & -1.52 & -0.402 & -0.174 & 0.355\\
$d_{\rm frac, out}^{H_2}$ & 0.0915 & 0.0815 & 0.0266 & 0.157 & 0.0210 & 0.00947\\
\hline
$\rho_{{\rm SF}, {H_{\rm I}}}$ & 0.563 & 0.564 & 0.530 & 0.596 & 0.562 & 0.564\\
$\rho_{{\rm SF}, H_2}$ & 0.893 & 0.894 & 0.876 & 0.911 & 0.899 & 0.907\\
$\rho_{{H_{\rm I}}, H_2}$ & 0.473 & 0.473 & 0.428 & 0.517 & 0.478 & 0.484\\
\hline
$\phi_{0}$ & 0.570 & 0.580 & 0.443 & 0.697 & 0.593 & 0.591\\
$d_{0}$ & -1.23 & -1.23 & -1.34 & -1.12 & -1.23 & -1.23\\
\hline
\end{tabular}\\[0.2cm]
\caption*{Primary model parameters constrained by the present analysis according to the fiducial model based on the representative xGASS / xCOLD GASS sample. The 26 model parameters can be organized into five groups. The first three groups contain the parameters for the marginal (1-dimensional) distributions of star formation rate (SFR), molecular, and neutral gas masses and their dependence on stellar mass. The fourth group list the correlation coefficients of the Gaussian copula. Finally, the parameters of the last group determine the fraction of galaxies belonging to the zero-component (i.e., those having vanishingly small SFRs and gas masses) and the dependence of this fraction on stellar mass. The name of each parameter is listed in the first column of the table. Columns 2-5 show the mean, median, and the 16th and 84th percentiles of their 1-dimensional probability distributions as provided by the Markov Chain Monte Carlo (MCMC) chain. The penultimate column lists the maximum a posteriori value obtained via optimization with the SLSQP minimizer\cite{Kraft1994} based on 51 starting values taken from the MCMC chain. Specifically, 50 randomly selected parameter combination are used in addition to the parameter vector from the MCMC chain with the largest probability density. The final column shows the maximum a posteriori (MAP) prediction when stellar mass errors are taken into account. The content of this table is available as Supplementary Data 1.}
\label{tab:ModelParameters}
\end{table}

\begin{table}
\centering
\caption{{\bf Parameters of the extended model}}
\linespread{1.22}
\footnotesize
\begin{tabular}{l|rrrrrr}
parameter & mean & median & 16th perc. & 84th perc. & MAP & MAP$^{\rm ME}$ \\
\hline
$\phi_{\rm scale}^{\rm SF}$ & 0.705 & 0.706 & 0.666 & 0.745 & 0.705 & 0.730\\
$d_{\rm scale}^{\rm SF}$ & -0.111 & -0.109 & -0.153 & -0.0690 & -0.0780 & -0.126\\
$\phi_{\rm shape}^{\rm SF}$ & -0.271 & -0.271 & -0.322 & -0.220 & -0.269 & -0.306\\
$d_{\rm shape}^{\rm SF}$ & -0.00462 & -0.00774 & -0.0615 & 0.0527 & -0.0591 & 0.0256\\
$d_{\rm scale, out}^{\rm SF}$ & 0.691 & 0.661 & 0.519 & 0.875 & 0.908 & 0.623\\
$d_{\rm shape, out}^{\rm SF}$ & -0.612 & -0.668 & -0.829 & -0.387 & -0.251 & -0.753\\
$d_{\rm frac, out}^{\rm SF}$ & 0.126 & 0.125 & 0.0681 & 0.180 & 0.0536 & 0.140\\
\hline
$\phi_{\rm scale}^{H_{\rm I}}$ & 0.336 & 0.337 & 0.295 & 0.378 & 0.332 & 0.339\\
$d_{\rm scale}^{H_{\rm I}}$ & 9.09 & 9.09 & 8.94 & 9.23 & 9.02 & 8.93\\
$\phi_{\rm shape}^{H_{\rm I}}$ & -0.00135 & -0.00117 & -0.0414 & 0.0386 & -0.00793 & -0.00791\\
$d_{\rm shape}^{H_{\rm I}}$ & -0.0842 & -0.0839 & -0.114 & -0.0542 & -0.0950 & -0.0813\\
$d_{\rm scale, out}^{H_{\rm I}}$ & 1.89 & 1.52 & 0.404 & 3.69 & 0.341 & 0.405\\
$d_{\rm shape, out}^{H_{\rm I}}$ & -0.928 & -1.17 & -1.80 & 0.286 & 0.466 & 0.449\\
$d_{\rm frac, out}^{H_{\rm I}}$ & 0.0282 & 0.00547 & 0.00107 & 0.0273 & 0.184 & 0.256\\
\hline
$\phi_{\rm scale}^{H_2}$ & 0.804 & 0.805 & 0.766 & 0.843 & 0.820 & 0.832\\
$d_{\rm scale}^{H_2}$ & 6.01 & 6.01 & 5.75 & 6.27 & 5.90 & 5.85\\
$\phi_{\rm shape}^{H_2}$ & -0.308 & -0.309 & -0.373 & -0.243 & -0.341 & -0.351\\
$d_{\rm shape}^{H_2}$ & 0.132 & 0.124 & 0.0653 & 0.202 & 0.151 & 0.115\\
$d_{\rm scale, out}^{H_2}$ & 0.906 & 0.690 & 0.403 & 1.34 & 0.453 & 0.744\\
$d_{\rm shape, out}^{H_2}$ & -1.05 & -1.11 & -1.65 & -0.458 & -1.13 & -0.716\\
$d_{\rm frac, out}^{H_2}$ & 0.0745 & 0.0642 & 0.0207 & 0.129 & 0.0840 & 0.0302\\
\hline
$\rho_{{\rm SF}, {H_{\rm I}}}$ & 0.601 & 0.602 & 0.569 & 0.634 & 0.601 & 0.598\\
$\rho_{{\rm SF}, H_2}$ & 0.880 & 0.881 & 0.864 & 0.897 & 0.884 & 0.888\\
$\rho_{{H_{\rm I}}, H_2}$ & 0.498 & 0.499 & 0.455 & 0.541 & 0.498 & 0.499\\
\hline
$\phi_{0}$ & 0.546 & 0.554 & 0.413 & 0.678 & 0.561 & 0.569\\
$d_{0}$ & -1.28 & -1.27 & -1.39 & -1.17 & -1.28 & -1.26\\
\hline
\end{tabular}\\[0.2cm]
\caption*{Primary model parameters constrained by the present analysis according to the extended model based on the extended xGASS / xCOLD GASS sample. The 26 model parameters can be organized into five groups. The first three groups contain the parameters for the marginal (1-dimensional) distributions of star formation rate (SFR), molecular, and neutral gas masses and their dependence on stellar mass. The fourth group list the correlation coefficients of the Gaussian copula. Finally, the parameters of the last group determine the fraction of galaxies belonging to the zero-component (i.e., those having vanishingly small SFRs and gas masses) and the dependence of this fraction on stellar mass. The name of each parameter is listed in the first column of the table. Columns 2-5 show the mean, median, and the 16th and 84th percentiles of their 1-dimensional probability distributions as provided by the Markov Chain Monte Carlo (MCMC) chain. The penultimate column lists the maximum a posteriori value obtained via optimization with the SLSQP minimizer\cite{Kraft1994} based on 51 starting values taken from the MCMC chain. Specifically, 50 randomly selected parameter combination are used in addition to the parameter vector from the MCMC chain with the largest probability density. The final column shows the maximum a posteriori (MAP) prediction when stellar mass errors are taken into account. The content of this table is available as Supplementary Data 2.}
\label{tab:ModelParametersExtended}
\end{table}

\begin{table}
\centering
\caption{{\bf Galaxy scaling relations in the fiducial model}}
\linespread{1.27}
\footnotesize
\begin{tabular}{c|rrrrrr}
parameter & mean & median & 16th perc. & 84th perc. & MAP & MAP$^{\rm ME}$ \\
\hline
$m^{\rm SFS}$ & 0.540 & 0.540 & 0.506 & 0.574 & 0.533 & 0.538\\
$m^{\rm NGS}$ & 0.331 & 0.331 & 0.299 & 0.362 & 0.325 & 0.318\\
$m^{\rm MGS}$ & 0.690 & 0.691 & 0.654 & 0.728 & 0.677 & 0.693\\
\hline
$n^{\rm SFS}$ & -0.173 & -0.172 & -0.195 & -0.150 & -0.172 & -0.171\\
$n^{\rm NGS}$ & 9.53 & 9.54 & 9.51 & 9.56 & 9.38 & 9.39\\
$n^{\rm MGS}$ & 8.78 & 8.78 & 8.75 & 8.80 & 8.75 & 8.77\\
\hline
$\Delta{}_{+,1}^{\rm SFS}$ & 0.378 & 0.379 & 0.359 & 0.396 & 0.372 & 0.381\\
$\Delta{}_{+,1}^{\rm NGS}$ & 0.402 & 0.402 & 0.390 & 0.413 & 0.402 & 0.402\\
$\Delta{}_{+,1}^{\rm MGS}$ & 0.313 & 0.315 & 0.285 & 0.339 & 0.335 & 0.334\\
\hline
$\Delta{}_{+,2}^{\rm SFS}$ & 0.331 & 0.332 & 0.316 & 0.345 & 0.326 & 0.334\\
$\Delta{}_{+,2}^{\rm NGS}$ & 0.349 & 0.349 & 0.341 & 0.358 & 0.350 & 0.350\\
$\Delta{}_{+,2}^{\rm MGS}$ & 0.280 & 0.282 & 0.257 & 0.301 & 0.298 & 0.297\\
\hline
$\Delta{}_{-,1}^{\rm SFS}$ & 0.531 & 0.532 & 0.494 & 0.568 & 0.519 & 0.538\\
$\Delta{}_{-,1}^{\rm NGS}$ & 0.580 & 0.579 & 0.556 & 0.604 & 0.580 & 0.580\\
$\Delta{}_{-,1}^{\rm MGS}$ & 0.412 & 0.416 & 0.364 & 0.458 & 0.450 & 0.449\\
\hline
$\Delta{}_{-,2}^{\rm SFS}$ & 0.657 & 0.658 & 0.602 & 0.710 & 0.639 & 0.666\\
$\Delta{}_{-,2}^{\rm NGS}$ & 0.729 & 0.728 & 0.692 & 0.766 & 0.730 & 0.730\\
$\Delta{}_{-,2}^{\rm MGS}$ & 0.487 & 0.491 & 0.421 & 0.550 & 0.539 & 0.537\\
\hline
\end{tabular}\\[0.2cm]
\caption*{The slope (first 3 rows), normalization (rows 4-6), and scatter (the remaining rows) of the star forming sequence (SFS), neutral gas sequence (NGS), and molecular gas sequence (MGS) as inferred by the fiducial model based on the representative xGASS / xCOLD GASS sample.  The first column lists the name of the derived model parameter, the other columns are defined analogously to those in Supplementary Table \ref{tab:ModelParameters}. Due to the involved non-linear mapping, the mean values of the derived parameters may differ (slightly) from the corresponding values calculated directly from the mean parameters listed in Supplementary Table \ref{tab:ModelParameters}.
The scatter is in general mass dependent. The values reported here correspond to galaxies with $M_{\rm star}=10^{10}$ $M_\odot$. Given the asymmetry of the SFS, NGS, and MGS, both upward and downward scatter are provided. For the SFS, the upward (downward) scatter $\Delta{}_{+, r}$ ($\Delta{}_{-, r}$) is defined as $1/r$ times the smallest increase (decrease) in $\lg{}{\rm SFR}$ that reduces the probability density of $\lg{}{\rm SFR}$ from its value at the peak of the SFS to $e^{-r^2/2}$ times the peak value. $\Delta{}_{+,r}=\Delta{}_{-,r}=\sigma$ for a normal distribution with a standard deviation of $\sigma$ for any chosen $r>0$. Upward and downward scatter are defined analogously for the neutral and molecular gas sequences. The content of this table is available as Supplementary Data 3.}
\label{tab:DerivedParameters}
\end{table}

\begin{table}
\centering
\caption{{\bf Galaxy scaling relations in the extended model}}
\linespread{1.27}
\footnotesize
\begin{tabular}{c|rrrrrr}
parameter & mean & median & 16th perc. & 84th perc. & MAP & MAP$^{\rm ME}$ \\
\hline
$m^{\rm SFS}$ & 0.576 & 0.575 & 0.539 & 0.612 & 0.576 & 0.579\\
$m^{\rm NGS}$ & 0.349 & 0.349 & 0.317 & 0.381 & 0.337 & 0.345\\
$m^{\rm MGS}$ & 0.722 & 0.723 & 0.687 & 0.759 & 0.717 & 0.731\\
\hline
$n^{\rm SFS}$ & -0.151 & -0.151 & -0.174 & -0.128 & -0.164 & -0.142\\
$n^{\rm NGS}$ & 9.55 & 9.56 & 9.54 & 9.58 & 9.44 & 9.38\\
$n^{\rm MGS}$ & 8.81 & 8.81 & 8.78 & 8.83 & 8.81 & 8.80\\
\hline
$\Delta{}_{+,1}^{\rm SFS}$ & 0.375 & 0.375 & 0.353 & 0.397 & 0.396 & 0.363\\
$\Delta{}_{+,1}^{\rm NGS}$ & 0.405 & 0.404 & 0.393 & 0.417 & 0.409 & 0.403\\
$\Delta{}_{+,1}^{\rm MGS}$ & 0.325 & 0.327 & 0.300 & 0.348 & 0.317 & 0.329\\
\hline
$\Delta{}_{+,2}^{\rm SFS}$ & 0.328 & 0.329 & 0.311 & 0.345 & 0.345 & 0.319\\
$\Delta{}_{+,2}^{\rm NGS}$ & 0.352 & 0.351 & 0.343 & 0.361 & 0.355 & 0.351\\
$\Delta{}_{+,2}^{\rm MGS}$ & 0.289 & 0.291 & 0.270 & 0.308 & 0.283 & 0.293\\
\hline
$\Delta{}_{-,1}^{\rm SFS}$ & 0.526 & 0.526 & 0.483 & 0.568 & 0.567 & 0.502\\
$\Delta{}_{-,1}^{\rm NGS}$ & 0.586 & 0.585 & 0.561 & 0.610 & 0.594 & 0.583\\
$\Delta{}_{-,1}^{\rm MGS}$ & 0.433 & 0.436 & 0.390 & 0.474 & 0.419 & 0.440\\
\hline
$\Delta{}_{-,2}^{\rm SFS}$ & 0.649 & 0.649 & 0.585 & 0.712 & 0.709 & 0.612\\
$\Delta{}_{-,2}^{\rm NGS}$ & 0.738 & 0.737 & 0.701 & 0.776 & 0.751 & 0.733\\
$\Delta{}_{-,2}^{\rm MGS}$ & 0.516 & 0.518 & 0.456 & 0.573 & 0.495 & 0.525\\
\hline
\end{tabular}\\[0.2cm]
\caption*{The slope (first 3 rows), normalization (rows 4-6), and scatter (the remaining rows) of the star forming sequence (SFS), neutral gas sequence (NGS), and molecular gas sequence (MGS) as inferred by the extended model based on the extended xGASS / xCOLD GASS sample.  The first column lists the name of the derived model parameter, the other columns are defined analogously to those in Supplementary Table \ref{tab:ModelParametersExtended}. Due to the involved non-linear mapping, the mean values of the derived parameters may differ (slightly) from the corresponding values calculated directly from the mean parameters listed in Supplementary Table \ref{tab:ModelParametersExtended}.
The scatter is in general mass dependent. The values reported here correspond to galaxies with $M_{\rm star}=10^{10}$ $M_\odot$. Given the asymmetry of the SFS, NGS, and MGS, both upward and downward scatter are provided. For the SFS, the upward (downward) scatter $\Delta{}_{+, r}$ ($\Delta{}_{-, r}$) is defined as $1/r$ times the smallest increase (decrease) in $\lg{}{\rm SFR}$ that reduces the probability density of $\lg{}{\rm SFR}$ from its value at the peak of the SFS to $e^{-r^2/2}$ times the peak value. $\Delta{}_{+,r}=\Delta{}_{-,r}=\sigma$ for a normal distribution with a standard deviation of $\sigma$ for any chosen $r>0$. Upward and downward scatter are defined analogously for the neutral and molecular gas sequences. The content of this table is available as Supplementary Data 4.}
\label{tab:DerivedParametersExtended}
\end{table}

\begin{figure}
\includegraphics[width=140mm]{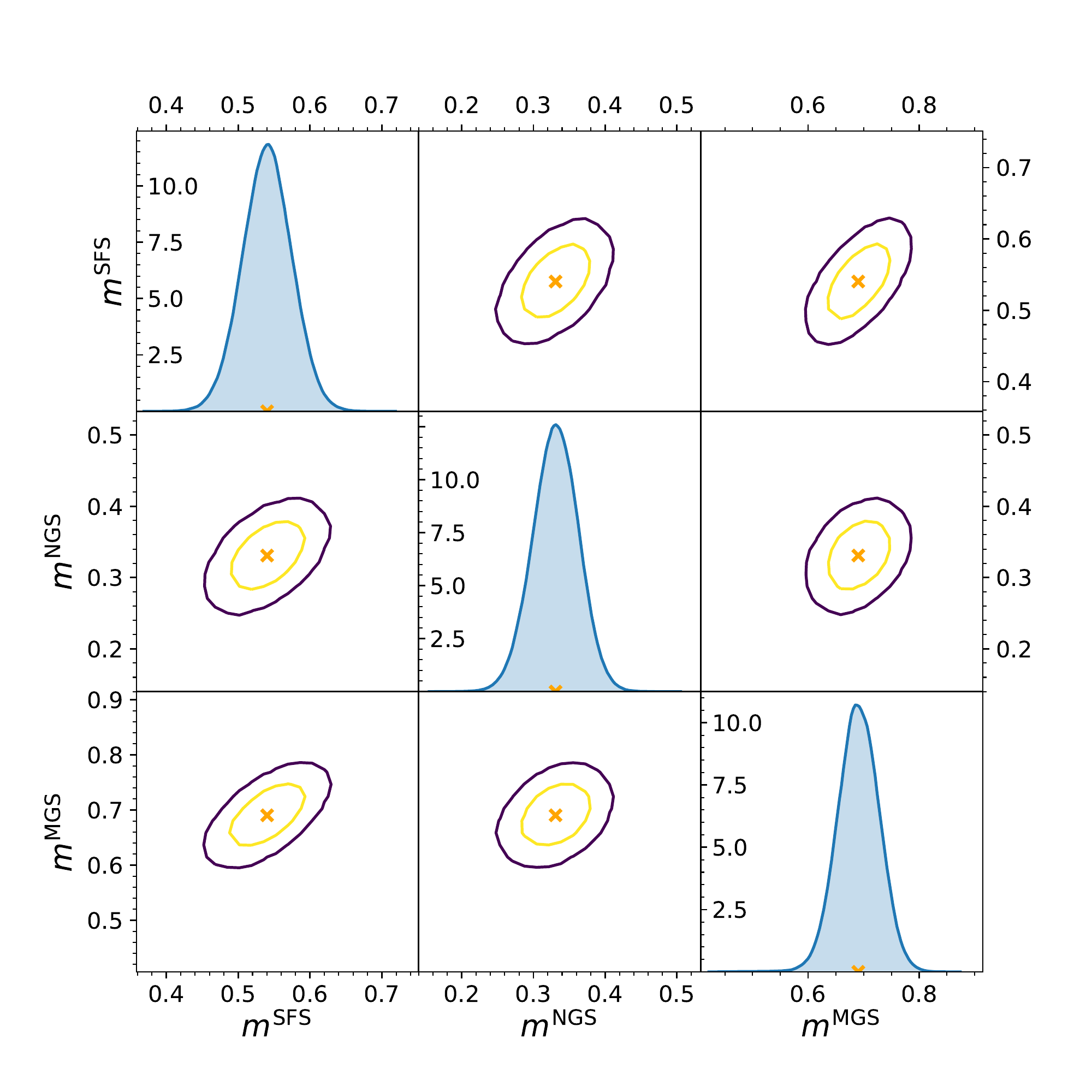}
\caption{{\bf Galaxy scaling relations -- Slopes.} The figure shows the modeling constraints for the slopes of the star forming sequence (SFS), neutral gas sequence (NGS), and molecular gas sequence (MGS): $m^{\rm SFS}$, $m^{\rm NGS}$, and $m^{\rm MGS}$ of nearby star-forming galaxies. Diagonal plots show the probability density of each parameter as provided by the Markov Chain Monte Carlo (MCMC) chain. Off-diagonal plots show probability density contours containing 68 and 95 percent of the probability. In each case, a cross indicates the average parameter value.}
\label{fig:ModelParametersSlope}
\end{figure}

\begin{figure}
\includegraphics[width=140mm]{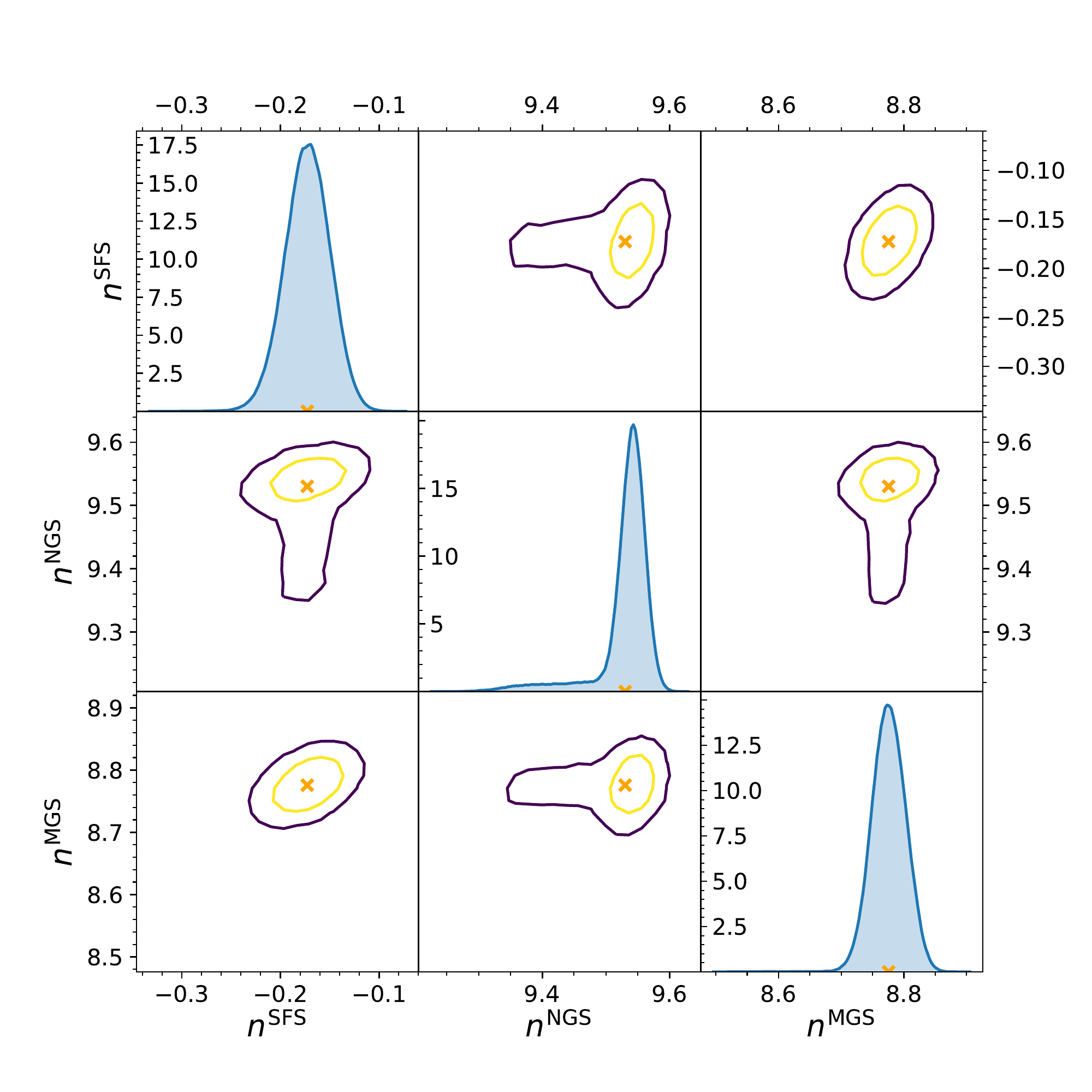}
\caption{{\bf Galaxy scaling relations -- Normalizations.} The figure shows the modeling constraints for the normalizations of the star forming sequence (SFS), neutral gas sequence (NGS), and molecular gas sequence (MGS) of nearby star-forming galaxies: $n^{\rm SFS}$, $n^{\rm NGS}$, and $n^{\rm MGS}$. Diagonal plots show the probability density of each parameter as provided by the Markov Chain Monte Carlo (MCMC) chain. Off-diagonal plots show probability density contours containing 68 and 95 percent of the probability. In each case, a cross indicates the average parameter value.}
\label{fig:ModelParametersNormalization}
\end{figure}

\begin{figure}
\includegraphics[width=140mm]{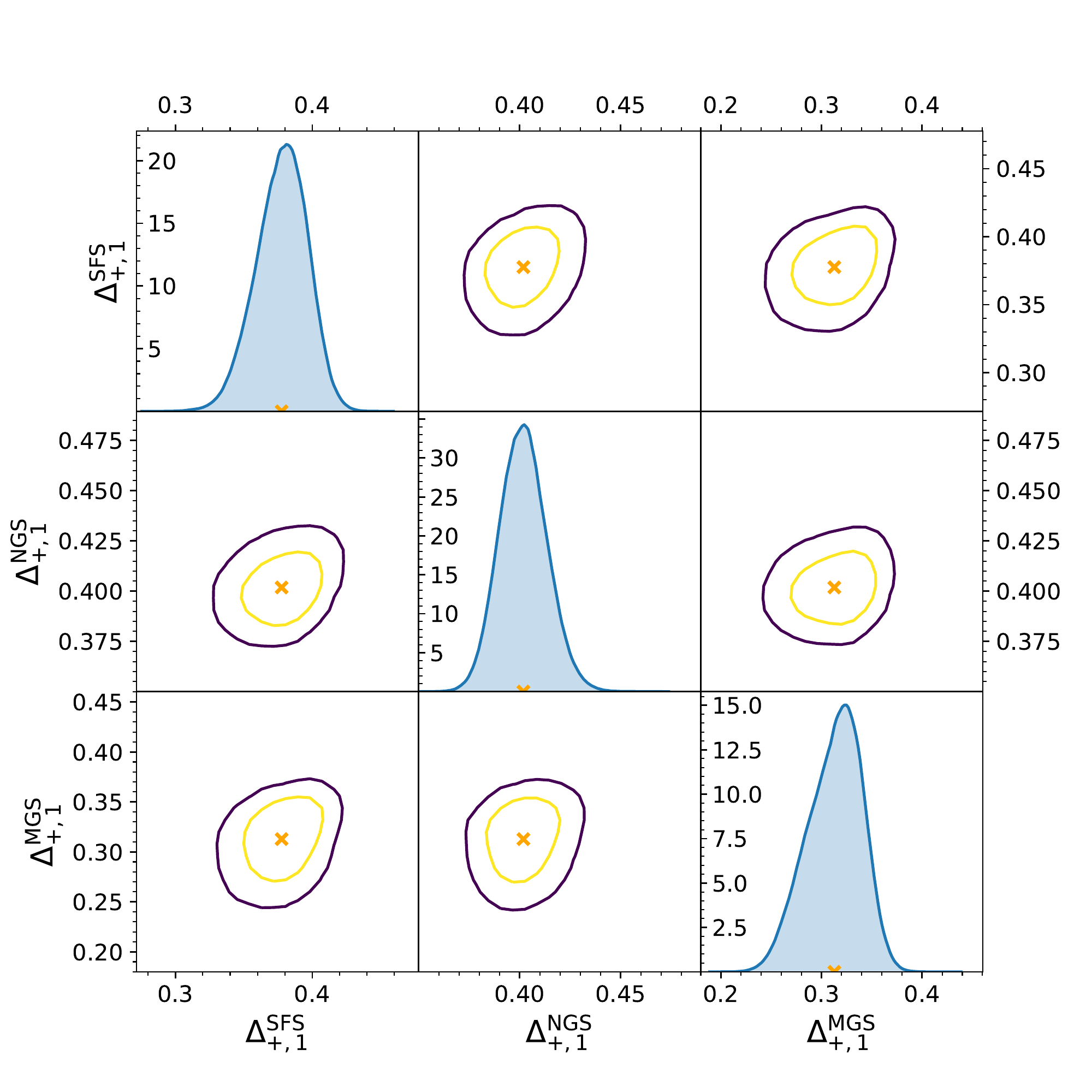}
\caption{{\bf Galaxy scaling relations -- Scatter.} The figure shows the modeling constraints for the upward scatter of the star forming sequence (SFS), neutral gas sequence (NGS), and molecular gas sequence (MGS) of nearby star-forming galaxies with $M_{\rm star}=10^{10}$ $M_\odot$: $\Delta{}_{+, 1}^{\rm SFS}$, $\Delta{}_{+, 1}^{\rm NGS}$, and $\Delta{}_{+, 1}^{\rm MGS}$. The scatter is calculated as described in the caption of Supplementary Table~\ref{tab:DerivedParameters} (for $r=1$). Diagonal plots show the probability density of each parameter as provided by the Markov Chain Monte Carlo (MCMC) chain. Off-diagonal plots show probability density contours containing 68 and 95 percent of the probability. In each case, a cross indicates the average parameter value.}
\label{fig:ModelParametersScatter}
\end{figure}

\begin{figure}
\includegraphics[width=140mm]{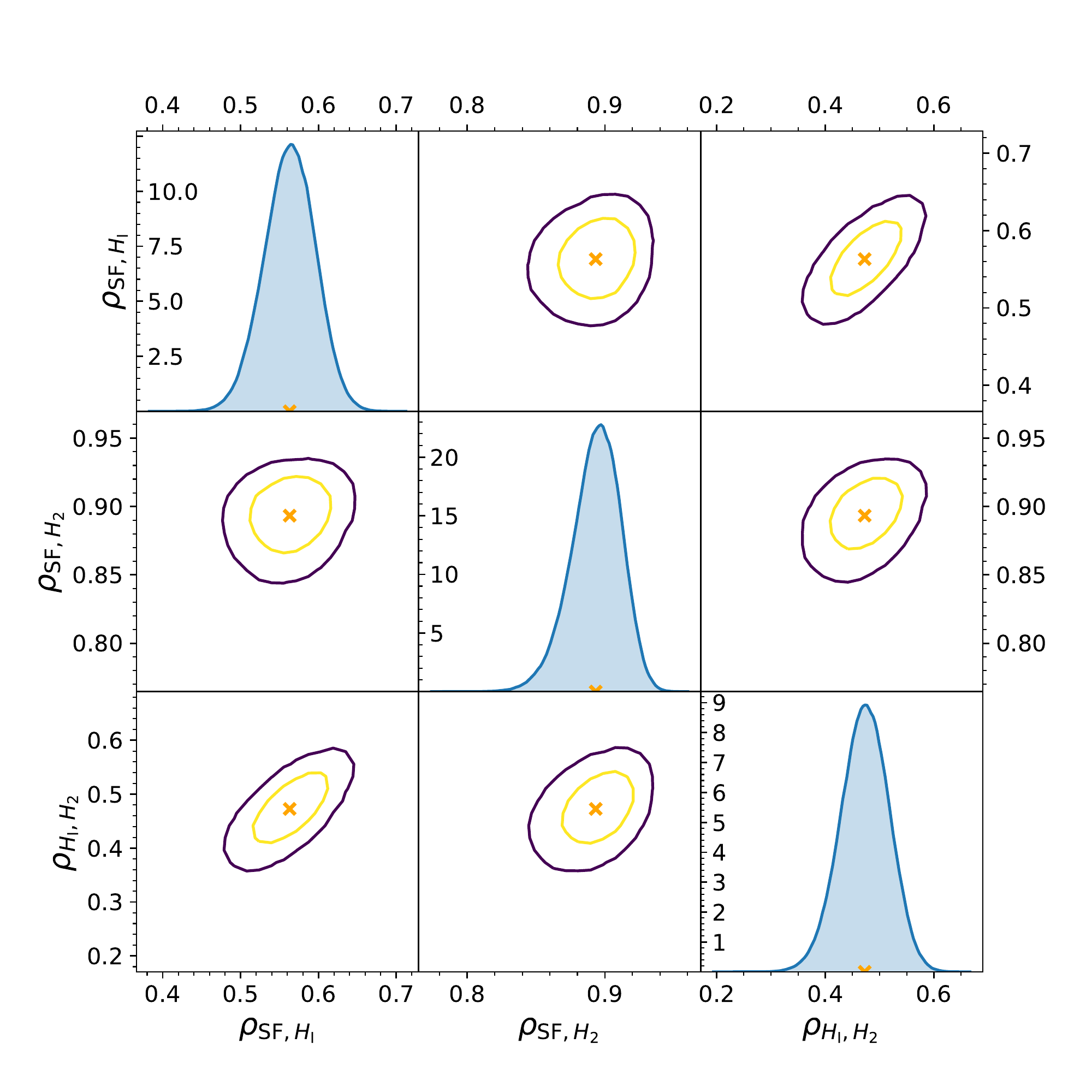}
\caption{{\bf Correlations between offsets from galaxy scaling relations.} The figure shows the modeling constraints for the off-diagonal entries of the correlation matrix $R$, see section \ref{sect:MultidimModel}. These parameters measure the correlation between SFRs, neutral, and molecular gas masses for galaxies of a given stellar mass and are correlation coefficients of the standardized variables $z_{\rm SF}, z_{\rm H_I}, z_{\rm H_2}$ introduced in section \ref{sect:Mock}. Diagonal plots show the probability density of each parameter as provided by the Markov Chain Monte Carlo (MCMC) chain. Off-diagonal plots show probability density contours containing 68 and 95 percent of the probability. In each case, a cross indicates the average parameter value. Offsets from the star forming sequence and the molecular gas sequence are tightly correlated as indicated by $\rho_{{\rm SF}, H_2}\sim{}0.9$.}
\label{fig:ModelParametersCorrelation}
\end{figure}

Supplementary Figures~\ref{fig:ModelParametersSlope}--\ref{fig:ModelParametersScatter} show the one- and two-dimensional probability densities of the slope, normalization, and upward scatter of the three sequences. In addition, Supplementary Figure~\ref{fig:ModelParametersCorrelation} shows the distribution of the correlation coefficients $\rho_{\rm SF, H_I}$, $\rho_{\rm SF, H_2}$, and $\rho_{\rm H_I, H_2}$.

\begin{figure}
\begin{center}
\includegraphics[width=160mm]{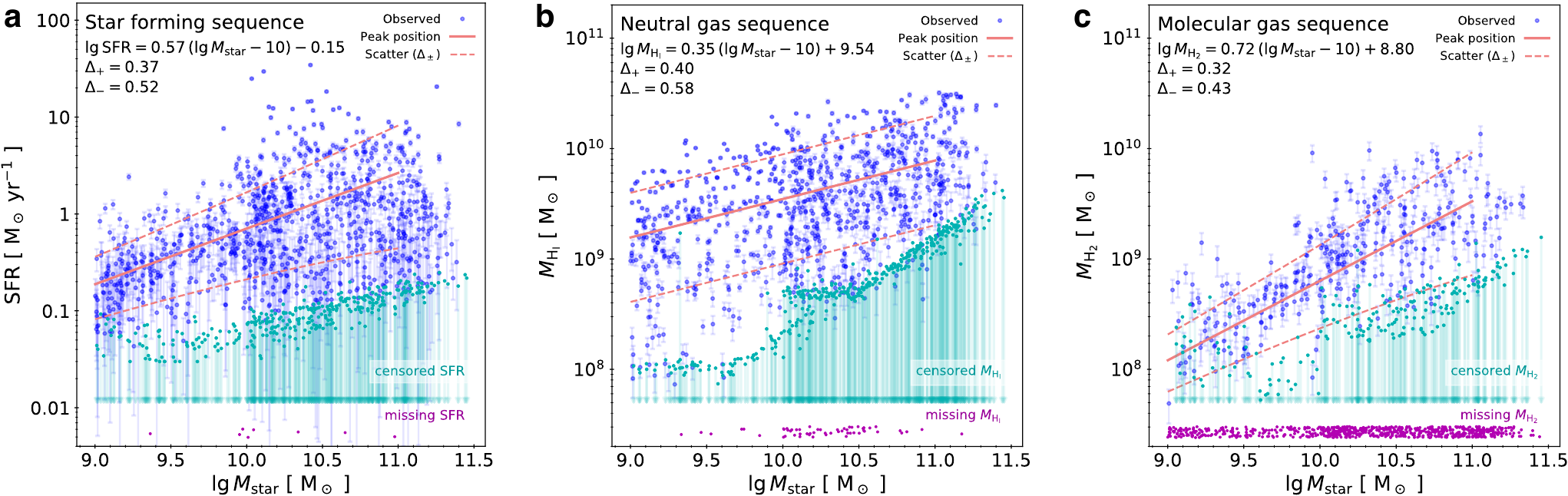}
\end{center}
\caption{{\bf Scaling relations of nearby galaxies in the extended model.}
Slope, normalization, and scatter of the star forming sequence ({\bf a}), neutral gas sequence ({\bf b}), and molecular gas sequence ({\bf c}). Points show the extended xGASS / xCOLD GASS sample\cite{Saintonge2017, Catinella2018}. Specifically, detected SFRs and gas masses are shown as blue circles with error bars indicating measurement uncertainties (one standard deviation). A large fraction of the observational data is either undetected/censored (cyan arrows) or missing (purple dots) necessitating careful modeling to avoid systematic biases. Peak position and scatter of each sequence, given as the average prediction of the extended model, are shown by solid and dashed lines. The peak position is defined as the mode of the conditional probability density of $\lg{}$SFR, $\lg{}M_{\rm H_I}$, and $\lg{}M_{\rm H_2}$ given $M_{\rm star}$. The predicted scaling of the peak position with stellar mass as well as the upward ($\Delta{}_+$) and downward ($\Delta{}_-$) scatter of each sequence for $M_{\rm star}=10^{10}$ $M_\odot$ galaxies are listed in the legend of each panel. The inferred galaxy scaling relations are similar to those shown in Fig.~\ref{fig:fig1} for the representative xGASS / xCOLD GASS sample.}
\label{fig:SequencesAlt}
\end{figure}

The figures and tables discussed above refer to the predictions of the fiducial model, i.e., they are based on the representative xGASS / xCOLD GASS sample. Instead, Supplementary Figure~\ref{fig:SequencesAlt} shows the model predictions when the extended xGASS / xCOLD GASS sample is used. The SFS, NGS, and MGS have a similar slope and scatter whether the representative (see Fig.~\ref{fig:fig1}) or the extended xGASS / xCOLD GASS sample is used.

\section*{Supplementary Note 2}
\label{sect:SuppSFP}

The orientation of the star forming plane (SFP) is calculated as follows. First, a random sample of $N_{\rm mock}=10^6$ actual data values ($\vec{y}$) is generated as described in section \ref{sect:Mock}. This sample (without the zero component) is used to calculate the probability density of $\Delta{}\lg{}M_{\rm H_2}$, $\Delta{}\lg{}M_{\rm H_I}$, $\Delta{}\lg{}{\rm SFR}$ with the help of a Gaussian kernel density estimate. Sample points with a probability density below the median value in the random sample are excluded to mitigate the larger uncertainties and higher leverage of sample points farther away from the mode of the probability distribution. Subsequently, a principal component analysis (PCA) is performed on the remaining sample points with the first and second principal component spanning the SFP. The third principal component is normal to the SFP and thus provides a convenient way to define the orientation of the SFP. Supplementary Table~\ref{tab:OrientationSFP} lists the orientation of the SFP for different $M_{\rm star}$ for the fiducial model showing only a weak mass dependence. Supplementary Table~\ref{tab:OrientationSFPAlt} shows the model predictions when the extended xGASS / xCOLD GASS sample is used instead.

\begin{table}
\centering
\caption{{\bf The orientation of the star forming plane in the fiducial model}}
\begin{tabular}{c|ccc}
$\lg{}M_{\rm star}/M_\odot$ & $n_{\rm H_2}$ & $n_{\rm H_I}$ & $n_{\rm SFR}$ \\
\hline
$9-9.1$ & 0.64 & -0.06 & -0.76 \\
$9.4-9.6$ & 0.66 & -0.07 & -0.75 \\
$9.9-10.1$ & 0.67 & -0.09 & -0.74 \\
$10.4-10.6$ & 0.68 & -0.09 & -0.73 \\
$10.9-11$ & 0.69 & -0.11 & -0.71 \\
$9 - 11$ & 0.67 & -0.08 & -0.74 \\
\end{tabular}\\[0.2cm]
\caption*{The stellar mass of galaxies is provided in the first column. Columns 2--4 list the components of a unit vector in $\Delta{}\lg{}M_{\rm H_2}$, $\Delta{}\lg{}M_{\rm H_I}$, $\Delta{}\lg{}{\rm SFR}$ space perpendicular to the star forming plane. This normal vector is calculated via principal component analysis based on mock samples generated by the fiducial model. The star forming plane is oriented almost perpendicular to the coordinate plane spanned by the  $M_{\rm H_2}$ -- SFR axes, highlighting the strong correlation between star formation rate (SFR) and molecular gas mass. The orientation of the star forming plane depends only weakly on stellar mass.}
\label{tab:OrientationSFP}
\end{table}

\begin{table}
\centering
\caption{{\bf The orientation of the star forming plane in the extended model}}
\begin{tabular}{c|ccc}
$\lg{}M_{\rm star}/M_\odot$ & $n_{\rm H_2}$ & $n_{\rm H_I}$ & $n_{\rm SFR}$ \\
\hline
9 & 0.67 & -0.09 & -0.74 \\
9.5 & 0.67 & -0.10 & -0.73 \\
10 & 0.68 & -0.11 & -0.73 \\
10.5 & 0.68 & -0.12 & -0.72 \\
11 & 0.68 & -0.14 & -0.72 \\
10 & 0.68 & -0.11 & -0.73 \\
\end{tabular}\\[0.2cm]
\caption*{The stellar mass of galaxies is provided in the first column. Columns 2--4 list the components of a unit vector in $\Delta{}\lg{}M_{\rm H_2}$, $\Delta{}\lg{}M_{\rm H_I}$, $\Delta{}\lg{}{\rm SFR}$ space perpendicular to the star forming plane. This normal vector is calculated via principal component analysis based on mock samples generated by the extended model. The star forming plane is oriented almost perpendicular to the coordinate plane spanned by the  $M_{\rm H_2}$ -- SFR axes, highlighting the strong correlation between star formation rate (SFR) and molecular gas mass. The orientation of the star forming plane depends only weakly on stellar mass.}
\label{tab:OrientationSFPAlt}
\end{table}

\section*{Supplementary Note 3}
\label{sect:SuppMolecularToNeutral}

\begin{figure}
\begin{tabular}{c}
\includegraphics[width=140mm]{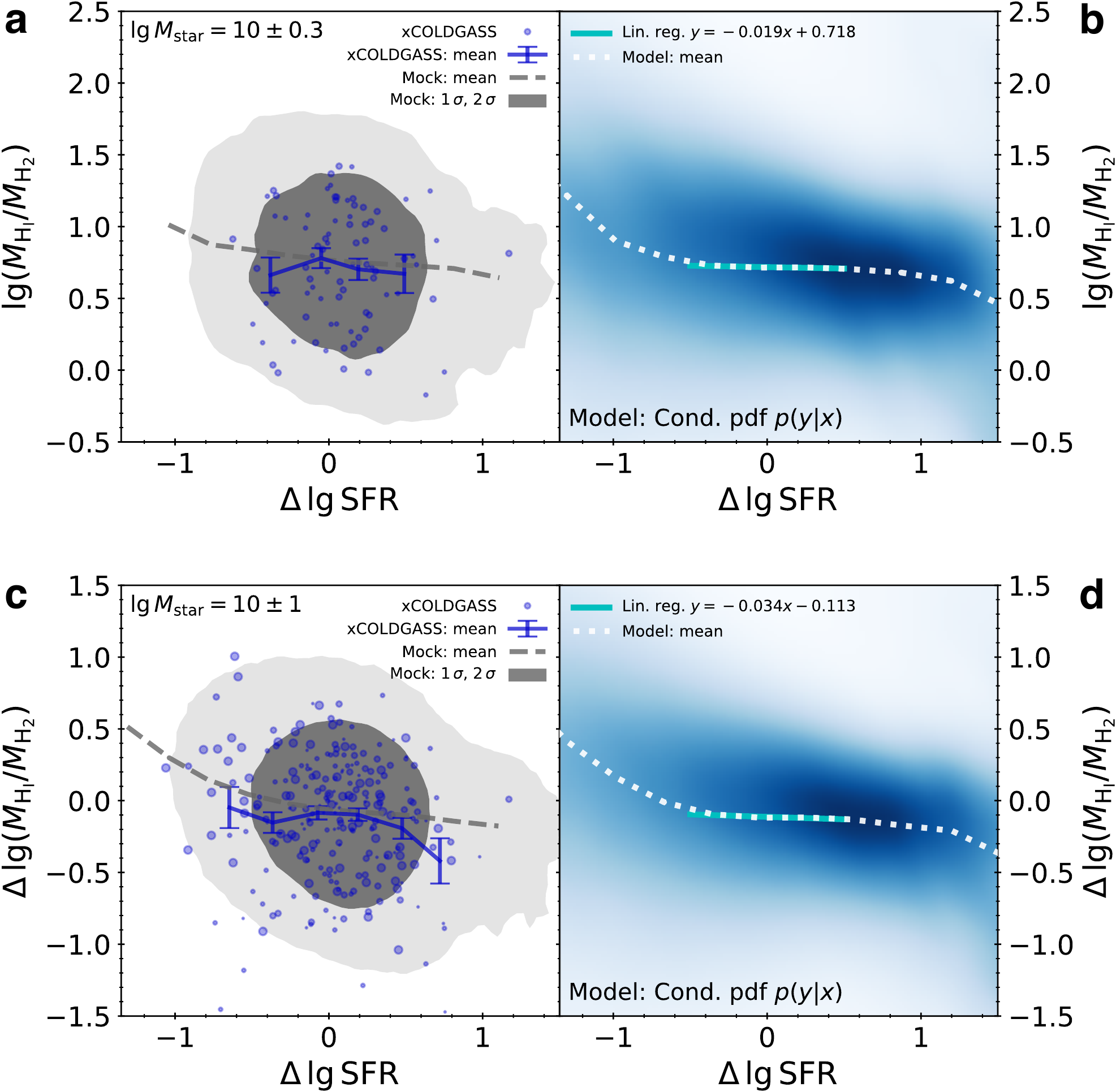}
\end{tabular}
\caption{{\bf Ratio between neutral and molecular gas mass.} {\bf a}, {\bf b} Neutral to molecular gas mass ratio for galaxies with $M_{\rm star}\sim{}10^{10}$ $M_\odot$ as function of the offset from the star forming sequence {\bf c}, {\bf d} Neutral to molecular gas mass ratio relative to its value at the peak of the neutral gas sequence (NGS) and molecular gas sequence (MGS), i.e., $\Delta{}\lg(M_{\rm H_I}/M_{\rm H_2}) = \lg(M_{\rm H_I}/M_{\rm H_2}) - \lg(M_{\rm H_I, NGS}/M_{\rm H_2, MGS})$. Here, all galaxies with $M_{\rm star}=10^{9-11}$ $M_\odot$ are included since the normalization removes most of the explicit stellar mass dependence.
{\bf a}, {\bf c} Observations and model-based mock data. Individual observations from the extended xGASS / xCOLD GASS sample are shown by points, while averages in bins of $\Delta{}\lg{}{\rm SFR}$ and the corresponding standard errors are shown by solid lines with error bars.  The shaded areas show $1\sigma{}$ and $2\sigma{}$ probability contours of the apparent $\lg(M_{\rm H_I}/M_{\rm H_2}) - \Delta{}\lg{}{\rm SFR}$ distribution as predicted by mock samples generated by the extended model. Mean values of $\lg(M_{\rm H_I}/M_{\rm H_2})$ for a given $\Delta{}\lg{}{\rm SFR}$ as predicted by the mock data are shown by a dashed line. 
{\bf b}, {\bf d} Model predictions. The blue shaded area is a map of the conditional probability density of the actual value of $\lg(M_{\rm H_I}/M_{\rm H_2})$ given $\Delta{}\lg{}{\rm SFR}$. Mean values are shown by a dotted line. A linear regression of the neutral-to-molecular gas mass ratio as function of $\Delta{}\lg{}{\rm SFR}\in{}[-0.5, 0.5]$ (solid line) shows a slope close to zero. Both $M_{\rm H_I}$ and $M_{\rm H_2}$ include a contribution from Helium.
}
\label{fig:NeutralToMolecular}
\end{figure}

Supplementary Figure~\ref{fig:NeutralToMolecular} analyzes how the neutral to molecular gas mass ratio scales with offset from the star forming sequence. For modest offsets ($\Delta{}\lg{}{\rm SFR}\in[-1,1]$) the scaling is weak, demonstrating that the neutral (and thus total gas) mass scales approximately with the molecular gas mass across the star forming sequence.

\section*{Supplementary Note 4}
\label{sect:SuppCatalog}

The joint xGASS / xCOLD GASS catalog is available as Supplementary Data 5.

\section*{Supplementary Note 5}
\label{sect:SuppDepletionTimes}

\begin{figure}
\includegraphics[width=140mm]{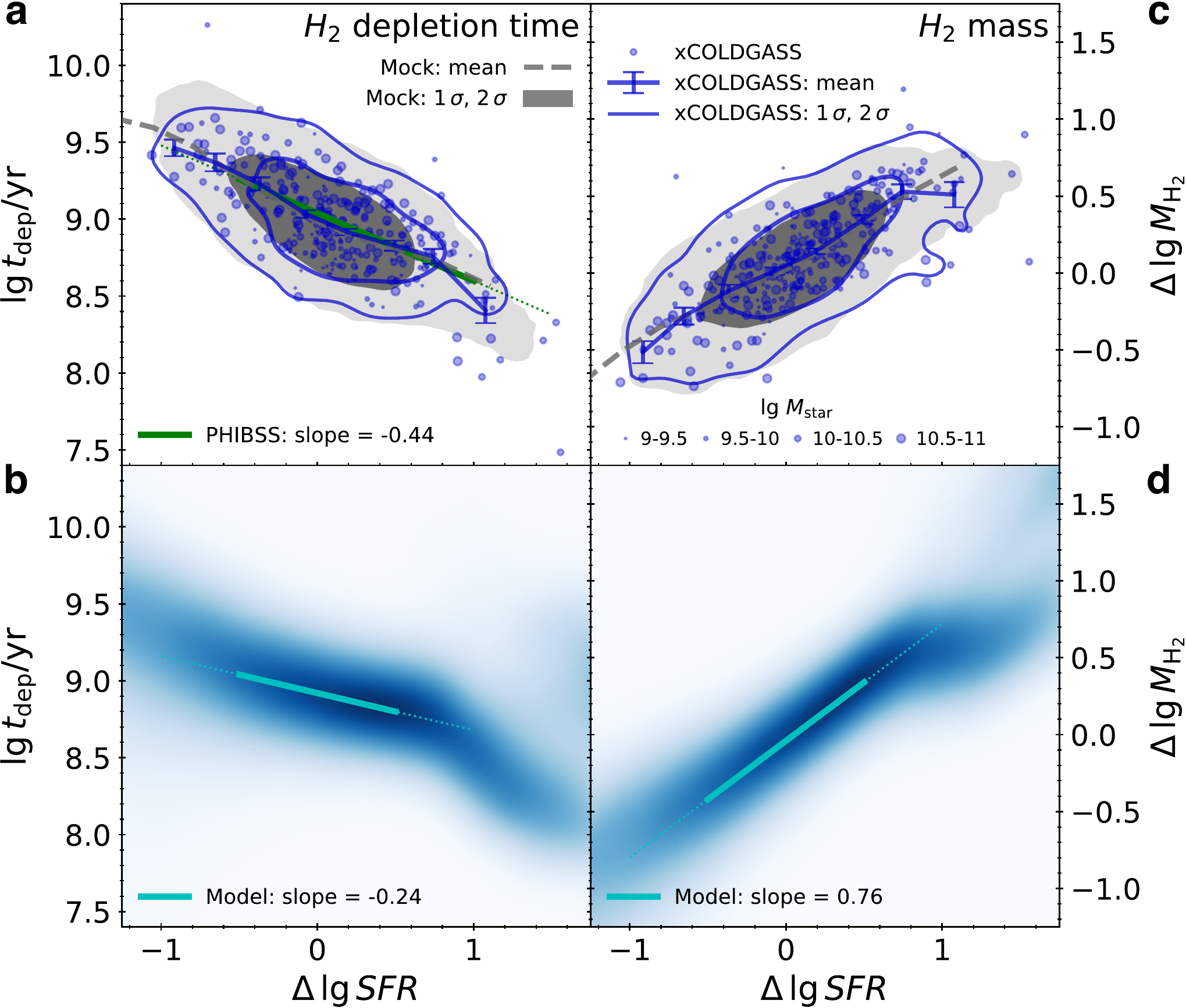}
\caption{{\bf Molecular gas depletion times and masses.} 
{\bf a}, {\bf b} Dependence of molecular gas depletion time on the offset from the star forming sequence. {\bf c}, {\bf d} Scaling of the offset from the molecular gas sequence with offset from the star forming sequence. {\bf a}, {\bf c} Observations and model-based mock data. Points and closed solid curves show individual values and contour lines for the extended xGASS / xCOLD GASS sample (see legend). In {\bf a} ({\bf c}), the shaded areas correspond to regions in $\lg{}t_{\rm dep}-\Delta{}\lg{}{\rm SFR}$ space ($\Delta{}\lg{}M_{\rm H_2}-\Delta{}\lg{}{\rm SFR}$ space) containing 68\% and 95\% of a mock sample generated from the model based on the extended xGASS / xCOLD GASS sample. $\Delta{}\lg{}M_{\rm H_2}$ is defined as $\lg{}M_{\rm H_2} - \lg{}M_{\rm H_2, MGS}$, i.e., analogously to $\Delta{}\lg{}{\rm SFR}$. Dashed lines (solid lines) show the mean value of the apparent $\lg{}t_{\rm dep}$ and $\Delta{}\lg{}M_{\rm H_2}$ distributions for a given $\Delta{}\lg{}{\rm SFR}$ in the mock sample (in the xGASS / xCOLD GASS sample). Error bars indicate standard errors of the mean. 
{\bf b}, {\bf d} Model predictions. The shaded area shows the actual conditional probability density (obtained from a kernel density estimate) of $\lg{}t_{\rm dep}$ ({\bf b}) and $\Delta{}\lg{}M_{\rm H_2}$ ({\bf d}) given $\Delta{}\lg{}{\rm SFR}$ for the model based on the extended xGASS / xCOLD GASS sample. The apparent gas depletion time scales more strongly with $\Delta{}\lg{}{\rm SFR}$ than the actual gas depletion time (slopes -0.44 vs -0.24), while the opposite holds true for the apparent and actual molecular gas masses (slopes -0.56 and -0.76).}
\label{fig:CondProb}
\end{figure}

Supplementary Figure~\ref{fig:CondProb} offers another look at how the molecular gas depletion time and molecular gas mass scale with offset from the star forming sequence, $\Delta{}\lg{}{\rm SFR}$. In contrast to Fig.~\ref{fig:fig4}, which presents average scaling relations, the panels of Supplementary Figure~\ref{fig:CondProb} show the distribution of the depletion time and gas mass as function of $\Delta{}\lg{}{\rm SFR}$. Mock data generated by the extended model reproduce well the distribution of observational data from xGASS / xCOLD GASS in $\lg{}t_{\rm dep}-\Delta{}\lg{\rm SFR}$ space. In agreement with the results shown in Fig.~\ref{fig:fig4}, the actual scaling of the molecular gas depletion time is significantly shallower than its apparent scaling for typical starforming (i.e., non-starbursting) galaxies.

\begin{figure}
\includegraphics[width=140mm]{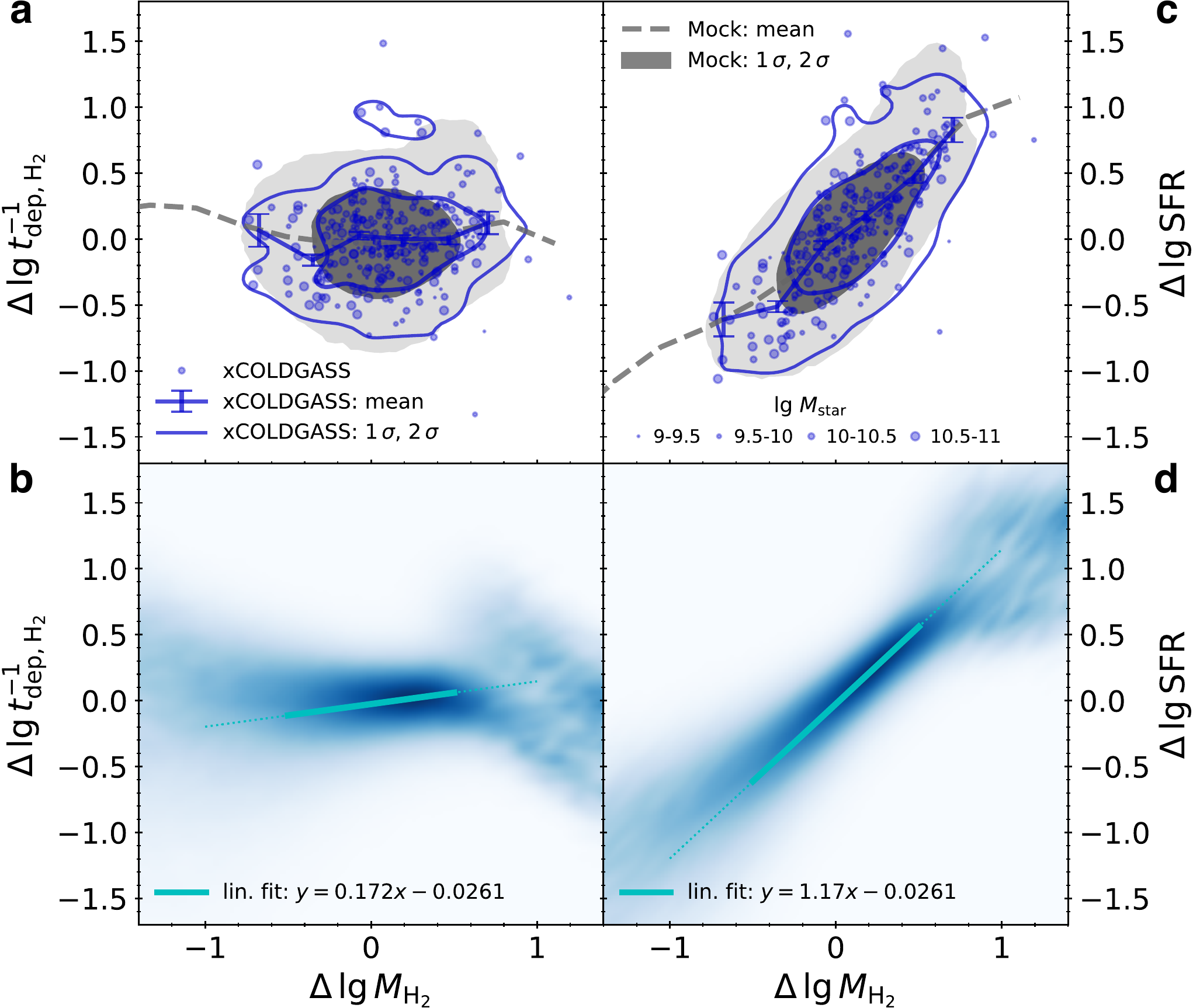}
\caption{{\bf Reciprocal molecular gas depletion times and star formation rates.} 
{\bf a}, {\bf b} Dependence of reciprocal molecular gas depletion time on the offset from the molecular gas sequence. {\bf c}, {\bf d} Scaling of the offset from the star forming sequence with offset from the molecular gas sequence.
{\bf a}, {\bf c} Observations and model-based mock data. Points and closed solid curves show individual values and contour lines for the extended xGASS / xCOLD GASS sample (see legend). In {\bf a} ({\bf c}), the shaded areas correspond to regions in $\Delta{}\lg{}t^{-1}_{\rm dep}-\Delta{}\lg{}M_{\rm H_2}$ space ($\Delta{}\lg{}{\rm SFR}-\Delta{}\lg{}M_{\rm H_2}$ space) containing 68\% and 95\% of a mock sample generated from the model based on the extended xGASS / xCOLD GASS sample. Here, $\Delta{}\lg{}t^{-1}_{\rm dep}$ is defined as $\Delta{}\lg{}{\rm SFR} - \Delta{}\lg{}M_{\rm H_2}$. Dashed lines (solid lines) show the mean value of the apparent $\Delta{}\lg{}t^{-1}_{\rm dep}$ and $\Delta{}\lg{}{\rm SFR}$ distributions for a given $\Delta{}\lg{}M_{\rm H_2}$ in the mock sample (in the xGASS / xCOLD GASS sample). Error bars indicate standard errors of the mean.
{\bf b}, {\bf d} Model predictions. The shaded area shows the actual conditional probability density (obtained from a kernel density estimate) of $\Delta{}\lg{}t^{-1}_{\rm dep}$ ({\bf b}) and $\Delta{}\lg{}{\rm SFR}$ ({\bf d}) given  $\Delta{}\lg{}M_{\rm H_2}$  for the model based on the extended xGASS / xCOLD GASS sample. The inverse molecular gas depletion time is only a weak function of $\Delta{}\lg{}M_{\rm H_2}$. In contrast, offsets from the star forming sequence and offsets from the molecular gas sequence are tightly correlated.}
\label{fig:CondProbMH2}
\end{figure}

Supplementary Figure~\ref{fig:CondProbMH2} is similar to Supplementary Figure~\ref{fig:CondProb} but plots the reciprocal molecular gas depletion time $t^{-1}_{\rm dep, H_2}(M_{\rm star}, M_{\rm H_2})$ as function of offset from the MGS. Here,
\[
\Delta{}\lg{}t^{-1}_{\rm dep}=\lg\left(\frac{{\rm SFR}}{M_{\rm H_2}}\right) - \lg\left(\frac{{\rm SFR}_{\rm SFS}(M_{\rm star})}{M_{\rm H_2, MGS}(M_{\rm star})}\right) = \Delta{}\lg{}{\rm SFR} - \Delta{}\lg{}M_{\rm H_2}
\]
Both the apparent and the actual scaling with $M_{\rm H_2}$ are very gradual, i.e., the $t^{-1}_{\rm dep, H_2}$ is only a very weak function of $M_{\rm H_2}$ at least for galaxies near the MGS.

\section*{Supplementary Discussion}
\label{sect:SuppDiscussion}

The following paragraphs analyze the slope of the SFS ($m$) and the slope of evolutionary tracks of galaxies in the $M_{\rm star}-{\rm SFR}$ plane ($\mu{}$) in the context of the evolutionary model (equations \ref{eq:ToyModel}, \ref{eq:EvModel}, and \ref{eq:EvModel2}). The slopes are defined as
\begin{equation}
m = \frac{\frac{\partial}{\partial{}s} \ln{}{\rm SFR}}{\frac{\partial}{\partial{}s} \ln{}M_{\rm star}}\,\textrm{  and   } \mu = \frac{\frac{\partial}{\partial{}t} \ln{}{\rm SFR}}{\frac{\partial}{\partial{}t} \ln{}M_{\rm star}}.
 \end{equation}
In general, $m\neq{}\mu$.

Starting from equation (\ref{eq:EvModel2}), I will simplify the math problem by assuming that $f_{\rm H_2}$ is a power-law function of $M_{\rm star}$ alone, i.e.,
$f_{\rm H_2}(M_{\rm star}) \propto{} M_{\rm star}^\gamma$. The power-law scaling could lead to an unphysical $f_{\rm H_2} > 1$ for galaxies with extreme masses. However, this assumption is adopted here merely to simplify the analytic derivation of the slopes $m$ and $\mu$. Equation (\ref{eq:EvModel2}) may still be solved numerically when a time dependent or non-power-law scaling is adopted. 
 
With the power-law scaling of $f_{\rm H_2}$, equation (\ref{eq:EvModel2}) simplifies to
\begin{equation}
\label{eq:EvModelSupp}
{\rm SFR}(t, s) = a\left[M_{\rm star}(t, s)\right]^{-\tilde{\beta}} \tilde{M}_{\rm gas}(t, s).
\end{equation}
with $\tilde{\beta} = (\beta-\gamma)/(1+\alpha)$ and $\tilde{M}_{\rm gas}(t, s)=\left[M_{\rm gas}(t, s)\right]^{1/(1+\alpha)}$. The exponents $\alpha$ and $\tilde{\beta}$ are assumed to be both larger than -1. Equation (\ref{eq:EvModelSupp})  can be solved by separation of variables for a given fixed $s$ and it has the solution
\begin{equation}
M_{\rm star}(t, s) = \left[a (1+\tilde{\beta}) \int_0^t \tilde{M}_{\rm gas}(t', s) dt'\right]^{\frac{1}{1+\tilde{\beta}}},
\end{equation}
and thus
\begin{equation}
\begin{split}
m(t, s) &= \frac{\frac{\partial}{\partial{}s}(-\tilde{\beta}\ln{}{\rm M_{\rm star}} + \ln{}\tilde{a} + \ln{}\tilde{M}_{\rm gas}) }  {\frac{\partial}{\partial{}s} \ln{}M_{\rm star}} = -\tilde{\beta} + \frac{\frac{\partial}{\partial{}s}\ln{}\tilde{M}_{\rm gas}}{\frac{\partial}{\partial{}s} \ln{}M_{\rm star}} \\
& =  -\tilde{\beta} + (1+\tilde{\beta}) \frac{\frac{\partial}{\partial{}s}\ln{}\tilde{M}_{\rm gas}(t, s)}{\frac{\partial}{\partial{}s} \ln{} \int_0^t \tilde{M}_{\rm gas}(t', s) dt'}.
\end{split}
\end{equation}

If the gas mass histories of galaxies on different tracks are scaled versions of each other, i.e., if $M_{\rm gas}(t, s)=g(t)h(s)$, then $\frac{\partial{}}{\partial{}s}\ln{} \tilde{M}_{\rm gas} = \frac{1}{1+\alpha}\frac{\partial{}}{\partial{}s} \ln{}h(s) = \frac{\partial{}}{\partial{}s}\ln{} \int_0^t \tilde{M}_{\rm gas}(t', s) dt'$. Thus, in this case $m=1$ and the slope of the SFS is exactly linear. This is true in particular if $M_{\rm gas}(t, s) = M_{\rm gas}(s)$.

Therefore, if equation (\ref{eq:EvModelSupp}) holds, a non-linear SFS implies that gas mass histories of different galaxies have different shapes. In particular, the sub-linear slope shown in Fig.~\ref{fig:fig6}b can be recovered by modeling the gas mass histories of galaxies in a way that more massive galaxies reach their maximum gas masses at earlier times and subsequently have faster declining gas masses, i.e., a form of `downsizing' \cite{PerezGonzalez2008a, Santini2014c}. Specifically, $M_{\rm gas}(t, s) \propto{} M_{\rm DM}(t, s) f_{\rm supp}(s, t) \frac{t}{\tau(s)}\exp(-t/\tau(s))$, where $M_{\rm DM}(t, s)$ is the average mass of the main progenitor of a $z=0$ dark matter halo of mass $s$\cite{Neistein2006b}, $\tau(s)$ is a (weak) power-law function of $s$, and $0\leq{}f_{\rm supp}(s, t)\leq{}1$ is a term that suppresses $M_{\rm gas}$ in low mass halos.

The slope of evolutionary tracks can be similarly calculated as
\begin{equation}
\mu(t, s) =-\tilde{\beta} + (1+\tilde{\beta}) \frac{\frac{\partial}{\partial{}t}\ln{}\tilde{M}_{\rm gas}(t, s)}{\frac{\partial}{\partial{}t} \ln{} \int_0^t \tilde{M}_{\rm gas}(t', s) dt'}  
 = -\tilde{\beta} + (1+\tilde{\beta}) \frac{\frac{\partial{}}{\partial{}t}\tilde{M}_{\rm gas}(t, s) \int_0^t \tilde{M}_{\rm gas}(t', s) dt'}{\tilde{M}_{\rm gas}(t, s)^2}.
\end{equation}
This equation shows that galaxies with constant gas masses evolve along tracks with a slope of $\mu=-\tilde{\beta}=(\gamma-\beta)/(1+\alpha)$. For instance, $\mu=0.11$ if $\alpha=-0.24$, $\beta=0.28$, $\gamma=0.69-0.33=0.36$, and $M_{\rm gas}$ is constant. The second term in the equation above is positive (negative) for galaxies with increasing (decreasing) gas masses. Hence, whether the slope of the evolutionary tracks of galaxies is greater or smaller than $-\tilde{\beta}$ is a direct measure of whether their gas masses increase or decrease.


\begin{thebibliography}{10}
\expandafter\ifx\csname url\endcsname\relax
  \def\url#1{\texttt{#1}}\fi
\expandafter\ifx\csname urlprefix\endcsname\relax\def\urlprefix{URL }\fi
\providecommand{\bibinfo}[2]{#2}
\providecommand{\eprint}[2][]{\url{#2}}

\bibitem{Krumholz2014c}
\bibinfo{author}{Krumholz, M.~R.}
\newblock \bibinfo{title}{{The big problems in star formation: The star
  formation rate, stellar clustering, and the initial mass function}}.
\newblock \emph{\bibinfo{journal}{Phys. Rep.}} \textbf{\bibinfo{volume}{539}},
  \bibinfo{pages}{49--134} (\bibinfo{year}{2014}).
\newblock \eprint{1402.0867}.

\bibitem{Noeske2007d}
\bibinfo{author}{Noeske, K.~G.} \emph{et~al.}
\newblock \bibinfo{title}{{Star Formation in AEGIS Field Galaxies since z =
  1.1: The Dominance of Gradually Declining Star Formation, and the Main
  Sequence of Star-forming Galaxies}}.
\newblock \emph{\bibinfo{journal}{Astrophys. J.}}
  \textbf{\bibinfo{volume}{660}}, \bibinfo{pages}{L43--L46}
  (\bibinfo{year}{2007}).

\bibitem{Daddi2007a}
\bibinfo{author}{Daddi, E.} \emph{et~al.}
\newblock \bibinfo{title}{{Multiwavelength Study of Massive Galaxies at
  $z\sim{}2$. I. Star Formation and Galaxy Growth}}.
\newblock \emph{\bibinfo{journal}{Astrophys. J.}}
  \textbf{\bibinfo{volume}{670}}, \bibinfo{pages}{156--172}
  (\bibinfo{year}{2007}).
\newblock \eprint{0705.2831}.

\bibitem{Dave2008}
\bibinfo{author}{Dav{\'{e}}, R.}
\newblock \bibinfo{title}{{The galaxy stellar mass-star formation rate
  relation: evidence for an evolving stellar initial mass function?}}
\newblock \emph{\bibinfo{journal}{Mon. Not. R. Astron. Soc.}}
  \textbf{\bibinfo{volume}{385}}, \bibinfo{pages}{147--160}
  (\bibinfo{year}{2008}).
\newblock \eprint{0710.0381}.

\bibitem{Lilly2013c}
\bibinfo{author}{Lilly, S.~J.}, \bibinfo{author}{Carollo, C.~M.},
  \bibinfo{author}{Pipino, A.}, \bibinfo{author}{Renzini, A.} \&
  \bibinfo{author}{Peng, Y.}
\newblock \bibinfo{title}{{GAS REGULATION OF GALAXIES: THE EVOLUTION OF THE
  COSMIC SPECIFIC STAR FORMATION RATE, THE METALLICITY-MASS-STAR-FORMATION RATE
  RELATION, AND THE STELLAR CONTENT OF HALOS}}.
\newblock \emph{\bibinfo{journal}{Astrophys. J.}}
  \textbf{\bibinfo{volume}{772}}, \bibinfo{pages}{119} (\bibinfo{year}{2013}).
\newblock \eprint{1303.5059}.

\bibitem{Feldmann2015}
\bibinfo{author}{Feldmann, R.} \& \bibinfo{author}{Mayer, L.}
\newblock \bibinfo{title}{{The Argo simulation - I. Quenching of massive
  galaxies at high redshift as a result of cosmological starvation}}.
\newblock \emph{\bibinfo{journal}{Mon. Not. R. Astron. Soc.}}
  \textbf{\bibinfo{volume}{446}}, \bibinfo{pages}{1939--1956}
  (\bibinfo{year}{2015}).
\newblock \eprint{1404.3212}.

\bibitem{Boselli2014a}
\bibinfo{author}{Boselli, a.} \emph{et~al.}
\newblock \bibinfo{title}{{Cold gas properties of the Herschel Reference
  Survey. II. Molecular and total gas scaling relations}}.
\newblock \emph{\bibinfo{journal}{Astron. Astrophys.}}
  \textbf{\bibinfo{volume}{564}}, \bibinfo{pages}{A66} (\bibinfo{year}{2014}).
\newblock \eprint{1401.8101}.

\bibitem{Genzel2015}
\bibinfo{author}{Genzel, R.} \emph{et~al.}
\newblock \bibinfo{title}{{Combined CO {\&} Dust Scaling Relations of Depletion
  Time and Molecular Gas Fractions with Cosmic Time, Specific Star Formation
  Rate and Stellar Mass}}.
\newblock \emph{\bibinfo{journal}{Astrophys. J.}}
  \textbf{\bibinfo{volume}{800}}, \bibinfo{pages}{20} (\bibinfo{year}{2014}).
\newblock \eprint{1409.1171}.

\bibitem{Saintonge2016}
\bibinfo{author}{Saintonge, A.} \emph{et~al.}
\newblock \bibinfo{title}{{Molecular and atomic gas along and across the main
  sequence of star-forming galaxies}}.
\newblock \emph{\bibinfo{journal}{Mon. Not. R. Astron. Soc.}}
  \textbf{\bibinfo{volume}{462}}, \bibinfo{pages}{1749--1756}
  (\bibinfo{year}{2016}).
\newblock \eprint{1607.05289}.

\bibitem{Tacconi2017}
\bibinfo{author}{Tacconi, L.~J.} \emph{et~al.}
\newblock \bibinfo{title}{{PHIBSS: Unified Scaling Relations of Gas Depletion
  Time and Molecular Gas Fractions}}.
\newblock \emph{\bibinfo{journal}{Astrophys. J.}}
  \textbf{\bibinfo{volume}{853}}, \bibinfo{pages}{179} (\bibinfo{year}{2018}).
\newblock \eprint{1702.01140}.

\bibitem{Bigiel2008}
\bibinfo{author}{Bigiel, F.} \emph{et~al.}
\newblock \bibinfo{title}{{THE STAR FORMATION LAW IN NEARBY GALAXIES ON SUB-KPC
  SCALES}}.
\newblock \emph{\bibinfo{journal}{Astron. J.}} \textbf{\bibinfo{volume}{136}},
  \bibinfo{pages}{2846--2871} (\bibinfo{year}{2008}).
\newblock \eprint{0810.2541}.

\bibitem{Krumholz2011a}
\bibinfo{author}{Krumholz, M.~R.}, \bibinfo{author}{Leroy, A.~K.} \&
  \bibinfo{author}{McKee, C.~F.}
\newblock \bibinfo{title}{{WHICH PHASE OF THE INTERSTELLAR MEDIUM CORRELATES
  WITH THE STAR FORMATION RATE?}}
\newblock \emph{\bibinfo{journal}{Astrophys. J.}}
  \textbf{\bibinfo{volume}{731}}, \bibinfo{pages}{25} (\bibinfo{year}{2011}).
\newblock \eprint{1101.1296}.

\bibitem{Saintonge2017}
\bibinfo{author}{Saintonge, A.} \emph{et~al.}
\newblock \bibinfo{title}{{xCOLD GASS: The Complete IRAM 30 m Legacy Survey of
  Molecular Gas for Galaxy Evolution Studies}}.
\newblock \emph{\bibinfo{journal}{Astrophys. J. Suppl. Ser.}}
  \textbf{\bibinfo{volume}{233}}, \bibinfo{pages}{22} (\bibinfo{year}{2017}).
\newblock \eprint{1710.02157}.

\bibitem{Catinella2018}
\bibinfo{author}{Catinella, B.} \emph{et~al.}
\newblock \bibinfo{title}{{xGASS: total cold gas scaling relations and
  molecular-to-atomic gas ratios of galaxies in the local Universe}}.
\newblock \emph{\bibinfo{journal}{Mon. Not. R. Astron. Soc.}}
  \textbf{\bibinfo{volume}{476}}, \bibinfo{pages}{875--895}
  (\bibinfo{year}{2018}).
\newblock \eprint{1802.02373}.

\bibitem{Saintonge2011g}
\bibinfo{author}{Saintonge, A.} \emph{et~al.}
\newblock \bibinfo{title}{{COLD GASS, an IRAM legacy survey of molecular gas in
  massive galaxies - II. The non-universality of the molecular gas depletion
  time-scale}}.
\newblock \emph{\bibinfo{journal}{Mon. Not. R. Astron. Soc.}}
  \textbf{\bibinfo{volume}{415}}, \bibinfo{pages}{61--76}
  (\bibinfo{year}{2011}).
\newblock \eprint{1104.0019}.

\bibitem{Shetty2013}
\bibinfo{author}{Shetty, R.}, \bibinfo{author}{Kelly, B.~C.} \&
  \bibinfo{author}{Bigiel, F.}
\newblock \bibinfo{title}{{Evidence for a non-universal Kennicutt-Schmidt
  relationship using hierarchical Bayesian linear regression}}.
\newblock \emph{\bibinfo{journal}{Mon. Not. R. Astron. Soc.}}
  \textbf{\bibinfo{volume}{430}}, \bibinfo{pages}{288--304}
  (\bibinfo{year}{2013}).
\newblock \eprint{1210.1218}.

\bibitem{Tacconi2020}
\bibinfo{author}{Tacconi, L.~J.}, \bibinfo{author}{Genzel, R.} \&
  \bibinfo{author}{Sternberg, A.}
\newblock \bibinfo{title}{{The Evolution of the Star-Forming Interstellar
  Medium Across Cosmic Time}}.
\newblock \emph{\bibinfo{journal}{Annu. Rev. Astron. Astrophys.}}
  \textbf{\bibinfo{volume}{58}}, \bibinfo{pages}{157--203}
  (\bibinfo{year}{2020}).
\newblock \eprint{2003.06245}.

\bibitem{Kelly2007}
\bibinfo{author}{Kelly, B.~C.}
\newblock \bibinfo{title}{{Some Aspects of Measurement Error in Linear
  Regression of Astronomical Data}}.
\newblock \emph{\bibinfo{journal}{Astrophys. J.}}
  \textbf{\bibinfo{volume}{665}}, \bibinfo{pages}{1489--1506}
  (\bibinfo{year}{2007}).

\bibitem{Robotham2015}
\bibinfo{author}{Robotham, A. S.~G.} \& \bibinfo{author}{Obreschkow, D.}
\newblock \bibinfo{title}{{Hyper-Fit: Fitting Linear Models to Multidimensional
  Data with Multivariate Gaussian Uncertainties}}.
\newblock \emph{\bibinfo{journal}{Publ. Astron. Soc. Aust.}}
  \textbf{\bibinfo{volume}{32}}, \bibinfo{pages}{e033} (\bibinfo{year}{2015}).

\bibitem{Feldmann2019a}
\bibinfo{author}{Feldmann, R.}
\newblock \bibinfo{title}{{LEO-Py: Estimating likelihoods for correlated,
  censored, and uncertain data with given marginal distributions}}.
\newblock \emph{\bibinfo{journal}{Astron. Comput.}}
  \textbf{\bibinfo{volume}{29}} (\bibinfo{year}{2019}).
\newblock \eprint{1910.02958}.

\bibitem{Feldmann2017}
\bibinfo{author}{Feldmann, R.}
\newblock \bibinfo{title}{{Are star formation rates of galaxies bimodal?}}
\newblock \emph{\bibinfo{journal}{Mon. Not. R. Astron. Soc. Lett.}}
  \textbf{\bibinfo{volume}{470}}, \bibinfo{pages}{L59--L63}
  (\bibinfo{year}{2017}).
\newblock \eprint{1705.03014}.

\bibitem{Donnari2019}
\bibinfo{author}{Donnari, M.} \emph{et~al.}
\newblock \bibinfo{title}{{The star formation activity of Illustris TNG
  galaxies: Main sequence, UVJ diagram, quenched fractions, and systematics}}.
\newblock \emph{\bibinfo{journal}{Mon. Not. R. Astron. Soc.}}
  \textbf{\bibinfo{volume}{485}}, \bibinfo{pages}{4817--4840}
  (\bibinfo{year}{2019}).

\bibitem{Sargent2012}
\bibinfo{author}{Sargent, M.~T.}, \bibinfo{author}{B{\'{e}}thermin, M.},
  \bibinfo{author}{Daddi, E.} \& \bibinfo{author}{Elbaz, D.}
\newblock \bibinfo{title}{{THE CONTRIBUTION OF STARBURSTS AND NORMAL GALAXIES
  TO INFRARED LUMINOSITY FUNCTIONS AT z {\textless} 2}}.
\newblock \emph{\bibinfo{journal}{Astrophys. J.}}
  \textbf{\bibinfo{volume}{747}}, \bibinfo{pages}{L31} (\bibinfo{year}{2012}).
\newblock \eprint{1202.0290}.

\bibitem{Foreman-Mackey2012a}
\bibinfo{author}{Foreman-Mackey, D.}, \bibinfo{author}{Hogg, D.~W.},
  \bibinfo{author}{Lang, D.} \& \bibinfo{author}{Goodman, J.}
\newblock \bibinfo{title}{{emcee: The MCMC Hammer}}.
\newblock \emph{\bibinfo{journal}{New York}} \bibinfo{pages}{1--22}
  (\bibinfo{year}{2012}).
\newblock \eprint{1202.3665}.

\bibitem{Cortese2011}
\bibinfo{author}{Cortese, L.}, \bibinfo{author}{Catinella, B.},
  \bibinfo{author}{Boissier, S.}, \bibinfo{author}{Boselli, A.} \&
  \bibinfo{author}{Heinis, S.}
\newblock \bibinfo{title}{{The effect of the environment on the Hi scaling
  relations}}.
\newblock \emph{\bibinfo{journal}{Mon. Not. R. Astron. Soc.}}
  \textbf{\bibinfo{volume}{415}}, \bibinfo{pages}{1797--1806}
  (\bibinfo{year}{2011}).
\newblock \eprint{1103.5889}.

\bibitem{Bahe2015a}
\bibinfo{author}{Bahe, Y.~M.} \& \bibinfo{author}{McCarthy, I.~G.}
\newblock \bibinfo{title}{{Star formation quenching in simulated group and
  cluster galaxies: when, how, and why?}}
\newblock \emph{\bibinfo{journal}{Mon. Not. R. Astron. Soc.}}
  \textbf{\bibinfo{volume}{447}}, \bibinfo{pages}{969--992}
  (\bibinfo{year}{2015}).
\newblock \eprint{1410.8161}.

\bibitem{Tacchella2016}
\bibinfo{author}{Tacchella, S.} \emph{et~al.}
\newblock \bibinfo{title}{{The confinement of star-forming galaxies into a main
  sequence through episodes of gas compaction, depletion and replenishment}}.
\newblock \emph{\bibinfo{journal}{Mon. Not. R. Astron. Soc.}}
  \textbf{\bibinfo{volume}{457}}, \bibinfo{pages}{2790--2813}
  (\bibinfo{year}{2016}).
\newblock \eprint{1509.02529}.

\bibitem{Feldmann2019}
\bibinfo{author}{Feldmann, R.}, \bibinfo{author}{Faucher-Gigu{\`{e}}re, C.-A.}
  \& \bibinfo{author}{Kere{\v{s}}, D.}
\newblock \bibinfo{title}{{The Galaxy-Halo Connection in Low-mass Halos}}.
\newblock \emph{\bibinfo{journal}{Astrophys. J.}}
  \textbf{\bibinfo{volume}{871}}, \bibinfo{pages}{L21} (\bibinfo{year}{2019}).
\newblock \eprint{1901.09039}.

\bibitem{Caplar2019a}
\bibinfo{author}{Caplar, N.} \& \bibinfo{author}{Tacchella, S.}
\newblock \bibinfo{title}{{Stochastic modelling of star-formation histories I:
  the scatter of the star-forming main sequence}}.
\newblock \emph{\bibinfo{journal}{Mon. Not. R. Astron. Soc.}}
  \textbf{\bibinfo{volume}{487}}, \bibinfo{pages}{3845--3869}
  (\bibinfo{year}{2019}).

\bibitem{Wang2019a}
\bibinfo{author}{Wang, E.}, \bibinfo{author}{Lilly, S.~J.},
  \bibinfo{author}{Pezzulli, G.} \& \bibinfo{author}{Matthee, J.}
\newblock \bibinfo{title}{{On the Elevation and Suppression of Star Formation
  within Galaxies}}.
\newblock \emph{\bibinfo{journal}{Astrophys. J.}}
  \textbf{\bibinfo{volume}{877}}, \bibinfo{pages}{132} (\bibinfo{year}{2019}).

\bibitem{Renzini2015}
\bibinfo{author}{Renzini, A.} \& \bibinfo{author}{Peng, Y.~J.}
\newblock \bibinfo{title}{{An objective definition for the main sequence of
  star-forming galaxies}}.
\newblock \emph{\bibinfo{journal}{Astrophys. J. Lett.}}
  \textbf{\bibinfo{volume}{801}}, \bibinfo{pages}{L29} (\bibinfo{year}{2015}).
\newblock \eprint{1502.01027}.

\bibitem{Speagle2014}
\bibinfo{author}{Speagle, J.~S.}, \bibinfo{author}{Steinhardt, C.~L.},
  \bibinfo{author}{Capak, P.~L.} \& \bibinfo{author}{Silverman, J.~D.}
\newblock \bibinfo{title}{{A HIGHLY CONSISTENT FRAMEWORK FOR THE EVOLUTION OF
  THE STAR-FORMING "MAIN SEQUENCE" FROM z ? 0-6}}.
\newblock \emph{\bibinfo{journal}{Astrophys. J. Suppl. Ser.}}
  \textbf{\bibinfo{volume}{214}}, \bibinfo{pages}{15} (\bibinfo{year}{2014}).
\newblock \eprint{1405.2041}.

\bibitem{Lorensen1987}
\bibinfo{author}{Lorensen, W.~E.} \& \bibinfo{author}{Cline, H.~E.}
\newblock \bibinfo{title}{{Marching cubes: A high resolution 3D surface
  construction algorithm}}.
\newblock In \emph{\bibinfo{booktitle}{Proc. 14th Annu. Conf. Comput. Graph.
  Interact. Tech. - SIGGRAPH '87}}, vol.~\bibinfo{volume}{21},
  \bibinfo{pages}{163--169} (\bibinfo{publisher}{ACM Press},
  \bibinfo{address}{New York, New York, USA}, \bibinfo{year}{1987}).

\bibitem{Janowiecki2018}
\bibinfo{author}{Janowiecki, S.}, \bibinfo{author}{Cortese, L.},
  \bibinfo{author}{Catinella, B.} \& \bibinfo{author}{Goodwin, A.~J.}
\newblock \bibinfo{title}{{Lurking systematics in predicting galaxy cold gas
  masses using dust luminosities and star formation rates}}.
\newblock \emph{\bibinfo{journal}{Mon. Not. R. Astron. Soc.}}
  \textbf{\bibinfo{volume}{476}}, \bibinfo{pages}{1390--1404}
  (\bibinfo{year}{2018}).
\newblock \eprint{1801.08687}.

\bibitem{Semenov2018}
\bibinfo{author}{Semenov, V.~A.}, \bibinfo{author}{Kravtsov, A.~V.} \&
  \bibinfo{author}{Gnedin, N.~Y.}
\newblock \bibinfo{title}{{How Galaxies Form Stars: The Connection between
  Local and Global Star Formation in Galaxy Simulations}}.
\newblock \emph{\bibinfo{journal}{Astrophys. J.}}
  \textbf{\bibinfo{volume}{861}}, \bibinfo{pages}{4} (\bibinfo{year}{2018}).
\newblock \eprint{1803.00007}.

\bibitem{Scoville2016}
\bibinfo{author}{Scoville, N.} \emph{et~al.}
\newblock \bibinfo{title}{{ISM MASSES AND THE STAR FORMATION LAW AT Z = 1 TO 6:
  ALMA OBSERVATIONS OF DUST CONTINUUM IN 145 GALAXIES IN THE COSMOS SURVEY
  FIELD}}.
\newblock \emph{\bibinfo{journal}{Astrophys. J.}}
  \textbf{\bibinfo{volume}{820}}, \bibinfo{pages}{83} (\bibinfo{year}{2016}).

\bibitem{Solomon1988}
\bibinfo{author}{Solomon, P.~M.} \& \bibinfo{author}{Sage, L.~J.}
\newblock \bibinfo{title}{{Star-formation rates, molecular clouds, and the
  origin of the far-infrared luminosity of isolated and interacting galaxies}}.
\newblock \emph{\bibinfo{journal}{Astrophys. J.}}
  \textbf{\bibinfo{volume}{334}}, \bibinfo{pages}{613} (\bibinfo{year}{1988}).

\bibitem{Whitaker2014b}
\bibinfo{author}{Whitaker, K.~E.} \emph{et~al.}
\newblock \bibinfo{title}{{CONSTRAINING THE LOW-MASS SLOPE OF THE STAR
  FORMATION SEQUENCE AT 0.5 {\textless} z {\textless} 2.5}}.
\newblock \emph{\bibinfo{journal}{Astrophys. J.}}
  \textbf{\bibinfo{volume}{795}}, \bibinfo{pages}{104} (\bibinfo{year}{2014}).

\bibitem{Krumholz2012}
\bibinfo{author}{Krumholz, M.~R.} \& \bibinfo{author}{Dekel, A.}
\newblock \bibinfo{title}{{METALLICITY-DEPENDENT QUENCHING OF STAR FORMATION AT
  HIGH REDSHIFT IN SMALL GALAXIES}}.
\newblock \emph{\bibinfo{journal}{Astrophys. J.}}
  \textbf{\bibinfo{volume}{753}}, \bibinfo{pages}{16} (\bibinfo{year}{2012}).

\bibitem{Mannucci2010}
\bibinfo{author}{Mannucci, F.}, \bibinfo{author}{Cresci, G.},
  \bibinfo{author}{Maiolino, R.}, \bibinfo{author}{Marconi, A.} \&
  \bibinfo{author}{Gnerucci, A.}
\newblock \bibinfo{title}{{A fundamental relation between mass, star formation
  rate and metallicity in local and high-redshift galaxies}}.
\newblock \emph{\bibinfo{journal}{Mon. Not. R. Astron. Soc.}}
  \textbf{\bibinfo{volume}{408}}, \bibinfo{pages}{2115--2127}
  (\bibinfo{year}{2010}).

\bibitem{Curti2020}
\bibinfo{author}{Curti, M.}, \bibinfo{author}{Mannucci, F.},
  \bibinfo{author}{Cresci, G.} \& \bibinfo{author}{Maiolino, R.}
\newblock \bibinfo{title}{{The mass-metallicity and the fundamental metallicity
  relation revisited on a fully Te-based abundance scale for galaxies}}.
\newblock \emph{\bibinfo{journal}{Mon. Not. R. Astron. Soc.}}
  \textbf{\bibinfo{volume}{491}}, \bibinfo{pages}{944--964}
  (\bibinfo{year}{2020}).
\newblock \eprint{1910.00597}.

\bibitem{Garnett2002}
\bibinfo{author}{Garnett, D.~R.}
\newblock \bibinfo{title}{{The Luminosity-Metallicity Relation, Effective
  Yields, and Metal Loss in Spiral and Irregular Galaxies}}.
\newblock \emph{\bibinfo{journal}{Astrophys. J.}}
  \textbf{\bibinfo{volume}{581}}, \bibinfo{pages}{1019--1031}
  (\bibinfo{year}{2002}).

\bibitem{Tremonti2004a}
\bibinfo{author}{Tremonti, C.~A.} \emph{et~al.}
\newblock \bibinfo{title}{{The Origin of the Mass-Metallicity Relation:
  Insights from 53,000 Star-forming Galaxies in the Sloan Digital Sky Survey}}.
\newblock \emph{\bibinfo{journal}{Astrophys. J.}}
  \textbf{\bibinfo{volume}{613}}, \bibinfo{pages}{898--913}
  (\bibinfo{year}{2004}).

\bibitem{Santini2014c}
\bibinfo{author}{Santini, P.} \emph{et~al.}
\newblock \bibinfo{title}{{The evolution of the dust and gas content in
  galaxies}}.
\newblock \emph{\bibinfo{journal}{Astron. Astrophys.}}
  \textbf{\bibinfo{volume}{562}}, \bibinfo{pages}{A30} (\bibinfo{year}{2014}).
\newblock \eprint{1311.3670}.

\bibitem{Suess2017}
\bibinfo{author}{Suess, K.~A.} \emph{et~al.}
\newblock \bibinfo{title}{{Massive quenched galaxies at z{\~{}}0.7 retain large
  molecular gas reservoirs}}.
\newblock \emph{\bibinfo{journal}{Astrophys. J. Lett.}}
  \textbf{\bibinfo{volume}{846}}, \bibinfo{pages}{L14} (\bibinfo{year}{2017}).
\newblock \eprint{1708.03337}.

\bibitem{Tomczak2016}
\bibinfo{author}{Tomczak, A.~R.} \emph{et~al.}
\newblock \bibinfo{title}{{THE SFR-M* RELATION AND EMPIRICAL STAR FORMATION
  HISTORIES FROM ZFOURGE AT 0.5 {\textless} z {\textless} 4}}.
\newblock \emph{\bibinfo{journal}{Astrophys. J.}}
  \textbf{\bibinfo{volume}{817}}, \bibinfo{pages}{118} (\bibinfo{year}{2016}).
\newblock \eprint{1510.06072}.

\bibitem{Schreiber2017}
\bibinfo{author}{Schreiber, C.} \emph{et~al.}
\newblock \bibinfo{title}{{The ALMA Redshift 4 Survey (AR4S)}}.
\newblock \emph{\bibinfo{journal}{Astron. Astrophys.}}
  \textbf{\bibinfo{volume}{599}}, \bibinfo{pages}{A134} (\bibinfo{year}{2017}).

\bibitem{Bouche2010}
\bibinfo{author}{Bouch{\'{e}}, N.} \emph{et~al.}
\newblock \bibinfo{title}{{THE IMPACT OF COLD GAS ACCRETION ABOVE A MASS FLOOR
  ON GALAXY SCALING RELATIONS}}.
\newblock \emph{\bibinfo{journal}{Astrophys. J.}}
  \textbf{\bibinfo{volume}{718}}, \bibinfo{pages}{1001--1018}
  (\bibinfo{year}{2010}).

\bibitem{Dave2012a}
\bibinfo{author}{Dav{\'{e}}, R.}, \bibinfo{author}{Finlator, K.} \&
  \bibinfo{author}{Oppenheimer, B.~D.}
\newblock \bibinfo{title}{{An analytic model for the evolution of the stellar,
  gas and metal content of galaxies}}.
\newblock \emph{\bibinfo{journal}{Mon. Not. R. Astron. Soc.}}
  \textbf{\bibinfo{volume}{421}}, \bibinfo{pages}{98--107}
  (\bibinfo{year}{2012}).
\newblock \eprint{1108.0426}.

\bibitem{Dekel2004}
\bibinfo{author}{Dekel, A.} \& \bibinfo{author}{Birnboim, Y.}
\newblock \bibinfo{title}{{Galaxy bimodality due to cold flows and shock
  heating}}.
\newblock \emph{\bibinfo{journal}{Mon. Not. R. Astron. Soc.}}
  \textbf{\bibinfo{volume}{368}}, \bibinfo{pages}{2--20}
  (\bibinfo{year}{2006}).
\newblock \eprint{0412300}.

\bibitem{Keres2005}
\bibinfo{author}{Kere{\v{s}}, D.}, \bibinfo{author}{Katz, N.},
  \bibinfo{author}{Weinberg, D.~H.} \& \bibinfo{author}{Dave, R.}
\newblock \bibinfo{title}{{How do galaxies get their gas?}}
\newblock \emph{\bibinfo{journal}{Mon. Not. R. Astron. Soc.}}
  \textbf{\bibinfo{volume}{363}}, \bibinfo{pages}{2--28}
  (\bibinfo{year}{2005}).
\newblock \eprint{0407095}.

\bibitem{Hopkins2014}
\bibinfo{author}{Hopkins, P.~F.} \emph{et~al.}
\newblock \bibinfo{title}{{Galaxies on FIRE (Feedback In Realistic
  Environments): Stellar Feedback Explains Cosmologically Inefficient Star
  Formation}}.
\newblock \emph{\bibinfo{journal}{Mon. Not. R. Astron. Soc.}}
  \textbf{\bibinfo{volume}{445}}, \bibinfo{pages}{581--603}
  (\bibinfo{year}{2014}).
\newblock \eprint{1311.2073}.

\bibitem{Vogelsberger2014}
\bibinfo{author}{Vogelsberger, M.} \emph{et~al.}
\newblock \bibinfo{title}{{Introducing the Illustris Project: simulating the
  coevolution of dark and visible matter in the Universe}}.
\newblock \emph{\bibinfo{journal}{Mon. Not. R. Astron. Soc.}}
  \textbf{\bibinfo{volume}{444}}, \bibinfo{pages}{1518--1547}
  (\bibinfo{year}{2014}).
\newblock \eprint{1405.2921}.

\bibitem{Schaye2015}
\bibinfo{author}{Schaye, J.} \emph{et~al.}
\newblock \bibinfo{title}{{The EAGLE project: simulating the evolution and
  assembly of galaxies and their environments}}.
\newblock \emph{\bibinfo{journal}{Mon. Not. R. Astron. Soc.}}
  \textbf{\bibinfo{volume}{446}}, \bibinfo{pages}{521--554}
  (\bibinfo{year}{2015}).
\newblock \eprint{1407.7040}.

\bibitem{Hobbs2020}
\bibinfo{author}{Hobbs, A.} \& \bibinfo{author}{Feldmann, R.}
\newblock \bibinfo{title}{{Positive feedback at the disc-halo interface}}.
\newblock \emph{\bibinfo{journal}{Mon. Not. R. Astron. Soc.}}
  \textbf{\bibinfo{volume}{498}}, \bibinfo{pages}{1140--1158}
  (\bibinfo{year}{2020}).
\newblock \eprint{2001.06012}.

\bibitem{Genzel2010}
\bibinfo{author}{Genzel, R.} \emph{et~al.}
\newblock \bibinfo{title}{{A study of the gas-star formation relation over
  cosmic time}}.
\newblock \emph{\bibinfo{journal}{Mon. Not. R. Astron. Soc.}}
  \textbf{\bibinfo{volume}{407}}, \bibinfo{pages}{2091--2108}
  (\bibinfo{year}{2010}).
\newblock \eprint{1003.5180}.

\bibitem{Hopkins2013c}
\bibinfo{author}{Hopkins, P.~F.} \emph{et~al.}
\newblock \bibinfo{title}{{Star formation in galaxy mergers with realistic
  models of stellar feedback and the interstellar medium}}.
\newblock \emph{\bibinfo{journal}{Mon. Not. R. Astron. Soc.}}
  \textbf{\bibinfo{volume}{430}}, \bibinfo{pages}{1901--1927}
  (\bibinfo{year}{2013}).
\newblock \eprint{1206.0011}.

\bibitem{Dekel2014}
\bibinfo{author}{Dekel, A.} \& \bibinfo{author}{Burkert, A.}
\newblock \bibinfo{title}{{Wet Disc Contraction to Galactic Blue Nuggets and
  Quenching to Red Nuggets}}.
\newblock \emph{\bibinfo{journal}{Mon. Not. R. Astron. Soc.}}
  \textbf{\bibinfo{volume}{438}}, \bibinfo{pages}{1870--1879}
  (\bibinfo{year}{2013}).
\newblock \eprint{1310.1074}.

\bibitem{Walter2016}
\bibinfo{author}{Walter, F.} \emph{et~al.}
\newblock \bibinfo{title}{{Alma Spectroscopic Survey in the Hubble Ultra Deep
  Field: Survey Description}}.
\newblock \emph{\bibinfo{journal}{Astrophys. J.}}
  \textbf{\bibinfo{volume}{833}}, \bibinfo{pages}{67} (\bibinfo{year}{2016}).
\newblock \eprint{1607.06768}.

\bibitem{Fevre2019}
\bibinfo{author}{{Le F{\`{e}}vre}, O.} \emph{et~al.}
\newblock \bibinfo{title}{{The ALPINE-ALMA [CII] survey}}.
\newblock \emph{\bibinfo{journal}{Astron. Astrophys.}}
  \textbf{\bibinfo{volume}{643}}, \bibinfo{pages}{A1} (\bibinfo{year}{2020}).
\newblock \eprint{1910.09517}.

\bibitem{Popping2019}
\bibinfo{author}{Popping, G.} \emph{et~al.}
\newblock \bibinfo{title}{{The ALMA Spectroscopic Survey in the HUDF: the
  Molecular Gas Content of Galaxies and Tensions with IllustrisTNG and the
  Santa Cruz SAM}}.
\newblock \emph{\bibinfo{journal}{Astrophys. J.}}
  \textbf{\bibinfo{volume}{882}}, \bibinfo{pages}{137} (\bibinfo{year}{2019}).
\newblock \eprint{1903.09158}.

\bibitem{Liang2019}
\bibinfo{author}{Liang, L.} \emph{et~al.}
\newblock \bibinfo{title}{{On the dust temperatures of high-redshift
  galaxies}}.
\newblock \emph{\bibinfo{journal}{Mon. Not. R. Astron. Soc.}}
  \textbf{\bibinfo{volume}{489}}, \bibinfo{pages}{1397--1422}
  (\bibinfo{year}{2019}).
\newblock \eprint{1902.10727}.

\bibitem{Mancuso2015}
\bibinfo{author}{Mancuso, C.} \emph{et~al.}
\newblock \bibinfo{title}{{PREDICTIONS for ULTRA-DEEP RADIO COUNTS of
  STAR-FORMING GALAXIES}}.
\newblock \emph{\bibinfo{journal}{Astrophys. J.}}
  \textbf{\bibinfo{volume}{810}}, \bibinfo{pages}{72} (\bibinfo{year}{2015}).

\bibitem{Blyth2015}
\bibinfo{author}{Blyth, S.} \emph{et~al.}
\newblock \bibinfo{title}{{Exploring Neutral Hydrogen and Galaxy Evolution with
  the SKA}}.
\newblock In \emph{\bibinfo{booktitle}{Proc. Adv. Astrophys. with Sq. Km. Array
  - PoS(AASKA14)}}, vol. \bibinfo{volume}{9-13-June-}, \bibinfo{pages}{128}
  (\bibinfo{publisher}{Sissa Medialab}, \bibinfo{address}{Trieste, Italy},
  \bibinfo{year}{2015}).
\newblock \eprint{1501.01295}.

\bibitem{Accurso2017}
\bibinfo{author}{Accurso, G.} \emph{et~al.}
\newblock \bibinfo{title}{{Deriving a multivariate $\alpha$CO conversion
  function using the [CII]/CO(1-0) ratio and its application to molecular gas
  scaling relations}}.
\newblock \emph{\bibinfo{journal}{Mon. Not. R. Astron. Soc.}}
  \textbf{\bibinfo{volume}{4766}}, \bibinfo{pages}{4750--4766}
  (\bibinfo{year}{2017}).
\newblock \eprint{1702.03888}.

\bibitem{Abazajian2009}
\bibinfo{author}{Abazajian, K.~N.} \emph{et~al.}
\newblock \bibinfo{title}{{The seventh data release of the sloan digital sky
  survey}}.
\newblock \emph{\bibinfo{journal}{Astrophys. Journal, Suppl. Ser.}}
  \textbf{\bibinfo{volume}{182}}, \bibinfo{pages}{543--558}
  (\bibinfo{year}{2009}).

\bibitem{Eales2017a}
\bibinfo{author}{Eales, S.} \emph{et~al.}
\newblock \bibinfo{title}{{The Galaxy End Sequence}}.
\newblock \emph{\bibinfo{journal}{Mon. Not. R. Astron. Soc.}}
  \textbf{\bibinfo{volume}{465}}, \bibinfo{pages}{3125--3133}
  (\bibinfo{year}{2017}).

\bibitem{Hogg2010}
\bibinfo{author}{Hogg, D.~W.}, \bibinfo{author}{Bovy, J.} \&
  \bibinfo{author}{Lang, D.}
\newblock \bibinfo{title}{{Data analysis recipes: Fitting a model to data}}
  (\bibinfo{year}{2010}).
\newblock \eprint{1008.4686}.

\bibitem{Baldry2006}
\bibinfo{author}{Baldry, I.~K.} \emph{et~al.}
\newblock \bibinfo{title}{{Galaxy bimodality versus stellar mass and
  environment}}.
\newblock \emph{\bibinfo{journal}{Mon. Not. R. Astron. Soc.}}
  \textbf{\bibinfo{volume}{373}}, \bibinfo{pages}{469--483}
  (\bibinfo{year}{2006}).

\bibitem{Schmidt1963}
\bibinfo{author}{Schmidt, M.}
\newblock \bibinfo{title}{{The Rate of Star Formation. II. The Rate of
  Formation of Stars of Different Mass.}}
\newblock \emph{\bibinfo{journal}{Astrophys. J.}}
  \textbf{\bibinfo{volume}{137}}, \bibinfo{pages}{758} (\bibinfo{year}{1963}).

\bibitem{Tinsley1980}
\bibinfo{author}{Tinsley, B.~M.}
\newblock \bibinfo{title}{{Evolution of the Stars and Gas in Galaxies}}.
\newblock \emph{\bibinfo{journal}{Fundam. Cosm. Phys.}}
  \textbf{\bibinfo{volume}{5}}, \bibinfo{pages}{287--388}
  (\bibinfo{year}{1980}).

\bibitem{Behroozi2018a}
\bibinfo{author}{Behroozi, P.}, \bibinfo{author}{Wechsler, R.~H.},
  \bibinfo{author}{Hearin, A.~P.} \& \bibinfo{author}{Conroy, C.}
\newblock \bibinfo{title}{{UniverseMachine: The correlation between galaxy
  growth and dark matter halo assembly from z = 0-10}}.
\newblock \emph{\bibinfo{journal}{Mon. Not. R. Astron. Soc.}}
  \textbf{\bibinfo{volume}{488}}, \bibinfo{pages}{3143--3194}
  (\bibinfo{year}{2019}).
\newblock \eprint{1806.07893}.

\bibitem{Kraft1994}
\bibinfo{author}{Kraft, D.}
\newblock \bibinfo{title}{{Algorithm 733; TOMP---Fortran modules for optimal
  control calculations}}.
\newblock \emph{\bibinfo{journal}{ACM Trans. Math. Softw.}}
  \textbf{\bibinfo{volume}{20}}, \bibinfo{pages}{262--281}
  (\bibinfo{year}{1994}).
\newblock
  \urlprefix\url{http://portal.acm.org/citation.cfm?doid=192115.192124}.

\bibitem{PerezGonzalez2008a}
\bibinfo{author}{P{\'{e}}rez-Gonz{\'{a}}lez, P.~G.} \emph{et~al.}
\newblock \bibinfo{title}{{The Stellar Mass Assembly of Galaxies from z = 0 to
  z = 4: Analysis of a Sample Selected in the Rest-Frame Near-Infrared with
  Spitzer}}.
\newblock \emph{\bibinfo{journal}{Astrophys. J.}}
  \textbf{\bibinfo{volume}{675}}, \bibinfo{pages}{234--261}
  (\bibinfo{year}{2008}).
\newblock \urlprefix\url{http://dx.doi.org/10.1086/523690}.
\newblock \eprint{0709.1354}.

\bibitem{Neistein2006b}
\bibinfo{author}{Neistein, E.}, \bibinfo{author}{van~den Bosch, F.~C.} \&
  \bibinfo{author}{Dekel, A.}
\newblock \bibinfo{title}{{Natural downsizing in hierarchical galaxy
  formation}}.
\newblock \emph{\bibinfo{journal}{Mon. Not. R. Astron. Soc.}}
  \textbf{\bibinfo{volume}{372}}, \bibinfo{pages}{933--948}
  (\bibinfo{year}{2006}).
\newblock \urlprefix\url{http://doi.wiley.com/10.1111/j.1365-2966.2006.10918.x}.

\end{thebibliography}
\end{document}